\documentclass[10pt,twocolumn,english,aps,manuscript,showpacs,showkeys,superscriptaddress,citeautoscript]{revtex4}
\usepackage[T1]{fontenc}
\usepackage[latin9]{inputenc}
\usepackage[active]{srcltx}
\usepackage{amsmath}
\usepackage{amssymb}
\usepackage{graphicx}

\makeatletter
\@ifundefined{textcolor}{}
{%
 \definecolor{BLACK}{gray}{0}
 \definecolor{WHITE}{gray}{1}
 \definecolor{RED}{rgb}{1,0,0}
 \definecolor{GREEN}{rgb}{0,1,0}
 \definecolor{BLUE}{rgb}{0,0,1}
 \definecolor{CYAN}{cmyk}{1,0,0,0}
 \definecolor{MAGENTA}{cmyk}{0,1,0,0}
 \definecolor{YELLOW}{cmyk}{0,0,1,0}
}

\include{shadowtext}

\usepackage{babel}

\usepackage{babel}

\usepackage{babel}

\usepackage{babel}

\makeatother

\usepackage{babel}
\begin{document}
\title{Universality and quasicritical exponents of one-dimensional
models displaying a quasitransition at finite temperatures}
\author{Onofre Rojas}
\address{Departamento de Fisica, Universidade Federal de Lavras, CP 3037, 37200000,
Lavras, MG, Brazil}
\author{Jozef Stre\v{c}ka}
\address{Department of Theoretical Physics and Astrophysics, Faculty of Science,
P.J. Šafárik University, Park Angelinum 9, 040 01, Košice, Slovak
Republic}
\author{Marcelo Leite Lyra}
\affiliation{Instituto de Física, Universidade Federal de Alagoas, 57072-970, Maceió,
AL, Brazil}
\author{Sergio Martins de Souza}
\affiliation{Departamento de Física, Universidade Federal de Lavras, CP 3037, 37200000,
Lavras, MG, Brazil}
\begin{abstract}
Quasicritical exponents of one-dimensional models displaying
a quasitransition at finite temperatures are examined in detail.
The quasitransition is characterized by intense sharp peaks in physical
quantities such as specific heat and magnetic susceptibility, which
are reminiscent of divergences accompanying a continuous (second-order)
phase transition. The question whether these robust finite peaks follow
some power law around the quasicritical temperature is addressed.
Although there is no actual divergence of these quantities at a quasicritical
temperature, a power-law behavior fits precisely both ascending as
well as descending part of the peaks in the vicinity but not too close
to a quasicritical temperature. The specific values of the quasicritical
exponents are rigorously calculated for a class of one-dimensional
models (e.g. Ising-XYZ diamond chain, coupled spin-electron double-tetrahedral
chain, Ising-XXZ two-leg ladder, and Ising-XXZ three-leg tube), whereas
the same set of quasicritical exponents implies a certain ``universality''
of quasitransitions of one-dimensional models. Specifically, the
values of the quasicritical exponents for one-dimensional models
are: $\alpha=\alpha'=3$ for the specific heat, $\gamma=\gamma'=3$
for the susceptibility and $\nu=\nu'=1$ for the correlation length. 
\end{abstract}
\maketitle

\section{Introduction}

Most one-dimensional systems in thermal equilibrium do not undergo
a phase-transition at finite temperatures. Several arguments have
been put forward giving support to the above statement as, for example,
the one based on the entropic contribution of domain walls by Landau
and Lifshitz \cite{landau}, the Perron-Frobenius theorem for the
non-degeneracy of the largest eigenvalue of a positive finite transfer
matrix \cite{perron}, and the van Hove's theorem stating that the
largest eigenvalue of a one-dimensional transfer matrix is an analytic
function \cite{hove}. A true phase-transition in one-dimensional
equilibrium systems may develop either when the model system depicts
long-range interactions or when a given interaction strength or a
local degree of freedom diverges \cite{kittel,weeks,dauxois}. Recently,
Sarkanych \textit{et al}. \cite{sarkanych} proposed an interesting
one-dimensional Potts model with \textquotedbl invisible states\textquotedbl{}
and short-range coupling. By term invisible, they refer to an additional
energy degeneracy, which contributes to the entropy, but not the interaction
energy.

In addition, Cuesta and Sanchez \cite{cuesta} summarized van Hove's
theorem is valid only under the following conditions: (i) the system
must be homogeneous, excluding automatically inhomogeneous systems,
i.e., disordered or aperiodic systems; (ii) the Hamiltonian does not
include particles position terms, such as, external fields; (iii)
the system must be considered as hard-core particles, while point-like
or soft particles may be excluded. Then, Cuesta and Sanchez \cite{cuesta}
generalized the non-existence theorem of phase transition at finite
temperatures. The extended theorem takes into account an external
field and point-like particles, which broadens the Van Hove's theorem,
although this is not yet a fully general theorem. For example, this
theorem cannot be applied for mixed particle chains or when more general
external fields are considered.

Recent exact calculations for a few paradigmatic models bear evidence
of remarkable \textquotedbl quasitransitions\textquotedbl{} of one-dimensional
lattice-statistical systems with short-range and non-singular interactions
\cite{gal15,tor16,roj16,str16}. In 2011 Timonin \cite{Timonin} introduced
the terms \textquotedbl pseudo-transitions\textquotedbl{} and \textquotedbl quasi-phases\textquotedbl{}
by investigating the Ising spin ice in a magnetic field when referring
to a sudden change in the first derivative and a sharp peak in the
second derivative of the free energy although there are neither true
discontinuities nor divergences in the appropriate derivatives of
the free energy. Although the physical property observed by Timonin
is precisely the same phenomenology presented by the models we study,
here we use just for convenience the term \textquotedbl quasi\textquotedbl{}
instead of \textquotedbl pseudo\textquotedbl . The quasitransitions
are thus reminiscent of discontinuous (first-order) phase transitions
due to abrupt temperature-driven changes of entropy, internal energy
and/or magnetization though these quantities display close to a quasicritical
temperature steep but continuous variations instead of real discontinuities
owing to analyticity of the free energy \cite{sou17}. On the other
hand, the quasitransitions of one-dimensional lattice-statistical
models are also reminiscent of continuous (second-order) phase transitions
due to massive rise of the correlation length, specific heat and susceptibility
in a vicinity of the quasicritical temperature though these quantities
exhibit very sharp and robust finite-size peaks instead of actual
divergences \cite{sou17}. The question whether these sizable peaks
follow some power-law behavior near the quasicritical temperature
is therefore quite intriguing and will be the main subject matter
of the present work. It will be verified that these physical quantities
indeed follow sufficiently close but not too close to a quasicritical
temperature power laws. In addition, it will be demonstrated that
the power-law behavior of seemingly diverse one-dimensional lattice-statistical
models can be described by a unique set of ``quasicritical'' exponents,
which enables us to conjecture the universality of ``quasitransitions''
of one-dimensional models.

A further investigation of quasitransitions and quasi-phases of one-dimensional
spin systems was considered in Ref. \cite{Isaac}, where the correlation
function around the quasitransition temperature was discussed. The
origin of quasitransition is however still not fully understood yet.
The residual entropy at zero temperature has been shown to be a good
indicator of the quasitransition as evidenced in Ref. \cite{org-psd}.

It is demonstrated that the observed quasicritical behavior is a
typical feature of a relatively wide class of one-dimensional Ising-Heisenberg
spin models and in this respect, it might be therefore of experimental
relevance for many real one-dimensional magnetic compounds of this
type (see for instance references \cite{S-HW,Heuvel-ch,Bel-Oh,Sahoo,S-Honda,Han-Strecka,Torr-Jmmm},
where Ising-Heisenberg models were applied to real compounds).

The present work is organized as follows. In Sec. 2 we will derive
analytic expressions for quasicritical exponents of the correlation
length, specific heat and magnetic susceptibility for one-dimensional
lattice-statistical models, which can be rigorously mapped onto the
effective Ising chain. In Sec. 3 we will specifically consider two
particular cases from this class of exactly solved one-dimensional
models: the spin-1/2 Ising-XYZ diamond chain and the coupled spin-electron
double-tetrahedral chain. In Sec. 4 we will further verify the universality
of quasicritical exponents by assuming another two exactly solved
one-dimensional lattice-statistical models falling beyond this class
of models: the spin-1/2 Ising-XXZ two-leg ladder and the spin-1/2
Ising-XXZ three-leg tube. Finally, our paper ends up with several
concluding remarks and future outlooks.

\section{quasicritical exponents}

\begin{figure}
\includegraphics[scale=0.3]{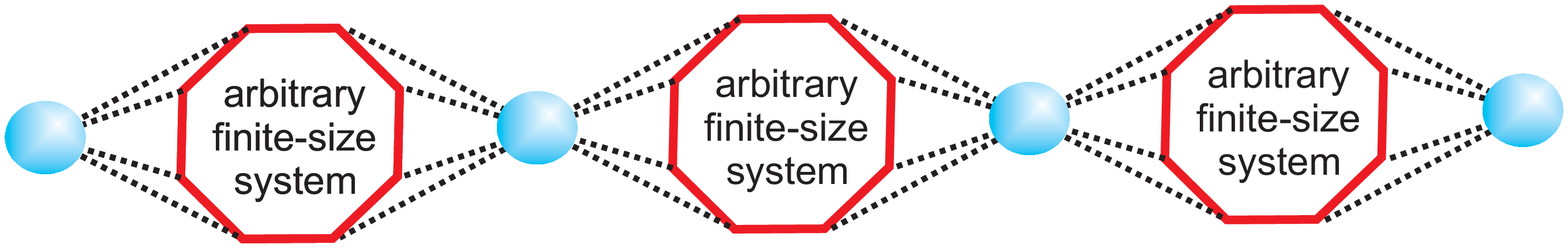} \caption{\label{fig0} A schematic representation of decorated one-dimensional
models, which consists of regularly alternating Ising spins (blue
balls) and arbitrary finite-size lattice-statistical system (red octagons).}
\end{figure}

It is firmly established that several one-dimensional models, which
can be viewed as the Ising chain decorated by arbitrary but finite
lattice-statistical system (see Fig. \ref{fig0} for a schematic representation),
are exactly tractable by taking advantage of a generalized decoration-iteration
transformation \cite{fis59,syo72,roj09,str10,strla,roj11}. The decoration-iteration
transformation furnishes a rigorous mapping correspondence between
the decorated one-dimensional models and the effective Ising chain.
This result would imply that the quasicritical exponents of the decorated
models can be obtained from the generic Ising chain given by the effective
Hamiltonian 
\begin{equation}
H=-\sum_{i=1}^{N}\left[K_{0}+Ks_{i}s_{i+1}+\frac{h_{{\rm eff}}}{2}(s_{i}+s_{i+1})\right],
\end{equation}
where $K_{0}$, $K$ and $h_{{\rm eff}}$ are effective temperature-dependent
parameters unambiguously given by the 'self-consistency' condition
of the decoration-iteration transformation \cite{fis59,syo72,roj09,str10,strla,roj11}.
By imposing the periodic boundary condition the effective Ising chain
can be readily solved by the transfer-matrix method, whereas the corresponding
transfer matrix can be generally expressed as follows \cite{sou17}
\begin{equation}
\mathbf{V}=\left(\begin{array}{cc}
w_{1} & w_{0}\\
w_{0} & w_{-1}
\end{array}\right).
\end{equation}
The Boltzmann factors pertinent to each sector (i.e. transfer-matrix
element) $n=\{-1,0,1\}$ are given by 
\begin{equation}
w_{n}=\sum_{k=0}g_{n,k}{\rm e}^{-\beta\varepsilon_{n,k}},\label{eq:wnk}
\end{equation}
where $\beta=1/(k_{B}T)$, $k_{B}$ is Boltzmann's constant, $T$
is the absolute temperature, $\varepsilon_{n,k}$ labels the energy
spectra for each sector $k=\{0,1,\ldots\}$ and $g_{n,k}$ denotes
the respective degeneracy of each energy level. It follows from the
transfer-matrix approach that the partition function can be expressed
in terms of transfer-matrix eigenvalues $\mathcal{Z}_{N}=\lambda_{+}^{N}+\lambda_{-}^{N}$,
which are explicitly given by 
\begin{equation}
\lambda_{\pm}=\tfrac{1}{2}\Bigl(w_{1}+w_{-1}\pm\sqrt{(w_{1}-w_{-1})^{2}+4w_{0}^{2}}\Bigr).\label{eq:L12}
\end{equation}
Then, the free energy attains in the thermodynamic limit ($N\rightarrow\infty$)
the following simple expression 
\begin{equation}
f=-\tfrac{1}{\beta}\ln\left[\tfrac{1}{2}\Bigl(w_{1}+w_{-1}+\sqrt{(w_{1}-w_{-1})^{2}+4w_{0}^{2}}\Bigr)\right].\label{eq:free-energ-1}
\end{equation}

Notice that all elements of the transfer matrix $\mathbf{V}$ are
strictly positive, except at zero temperature. Therefore its eigenvalues
are distinct and analytical according to Eq. \eqref{eq:L12}, in agreement
with the Perron-Frobenius theorem for matrices with all positive matrix
elements. This implies in the absence of a true finite-temperature
phase-transition in the one-dimensional Ising model.

A crossing of the transfer matrix eigenvalues would be required to
achieve non-analiticity of the free-energy as it is expected in a
phase-transition. It has been recently argued \cite{sou17} that a
quasitransition may occur when the following condition is satisfied:
\begin{equation}
|w_{1}-w_{-1}|>w_{0}\gtrsim0.\label{cc}
\end{equation}
which can be reached at finite temperatures in a large class of effectively
one-dimensional model systems. In what follows, we will unveil the
leading behavior of some typical thermodynamic quantities under the
above condition. For further convenience, it is therefore useful to
define the small-size parameter $\bar{w}_{0}=\frac{w_{0}}{|w_{1}-w_{-1}|}\rightarrow0$,
which is suitable for Taylor series expansion. At first, let us consider
the particular case when $w_{1}>w_{-1}$, then, the free energy \eqref{eq:free-energ-1}
becomes 
\begin{alignat}{1}
f= & -\tfrac{1}{\beta}\ln\left[w_{1}+\tfrac{1}{2}\left(w_{1}-w_{-1}\right)\Bigl(\sqrt{1+4\bar{w}_{0}^{2}}-1\Bigr)\right],\nonumber \\
= & -\tfrac{1}{\beta}\ln(w_{1})-\tfrac{1}{\beta}\ln\left[1+\tfrac{\left(w_{1}-w_{-1}\right)(\sqrt{1+4\bar{w}_{0}^{2}}-1)}{2w_{1}}\right].\label{eq:free-energ-2}
\end{alignat}
The last term of the second logarithm satisfies the following condition
\begin{equation}
0<\tfrac{\left(w_{1}-w_{-1}\right)(\sqrt{1+4\bar{w}_{0}^{2}}-1)}{2w_{1}}<1
\end{equation}
and this condition guarantees convergence of the Taylor series expansion
around $\bar{w}_{0}=0$. Hence, the first term will be more relevant
than the higher-order contributions arising from the Taylor series
expansion $\bar{w}_{0}\rightarrow0$. Analogously, the similar expression
can be obtained for the other particular case $w_{1}<w_{-1}$ by a
mere inter-change of $w_{1}\leftrightarrow w_{-1}$. To summarize,
the free energy \eqref{eq:free-energ-1} can be recast using the Taylor
series expansion around $\bar{w}_{0}\rightarrow0$ to the following
form 
\begin{equation}
f=\begin{cases}
-\tfrac{1}{\beta}\ln(w_{1})-\tfrac{1}{\beta}\tfrac{w_{0}^{2}}{w_{1}\left(w_{1}-w_{-1}\right)}+\mathcal{O}(\bar{w}_{0}^{3}), & w_{1}>w_{-1}\\
-\tfrac{1}{\beta}\ln(\tilde{w}_{1}+\tilde{w}_{0}), & w_{1}=w_{-1}\\
-\tfrac{1}{\beta}\ln(w_{-1})-\tfrac{1}{\beta}\tfrac{w_{0}^{2}}{w_{-1}\left(w_{-1}-w_{1}\right)}+\mathcal{O}(\bar{w}_{0}^{3}), & w_{1}<w_{-1}
\end{cases}\label{f-apx}
\end{equation}
where $\tilde{w}_{1}=w_{1}=w_{-1}$, and $w_{0}=\tilde{w}_{0}$ under
the specific condition $w_{1}=w_{-1}$. It is important to stress
that the additional condition $|w_{1}-w_{-1}|\gg w_{0}$ must the
fulfilled for the validity of the above asymptotic expansions when
$w_{1}\neq w_{-1}$.

In order to characterize the power-law behavior emergent close to
the quasitransition, it is useful to rewrite the Boltzmann factor
in terms of the relative difference $\tau=\left(T_{p}-T\right)/T_{p}$
between temperature $T$ and quasicritical temperature $T_{p}$,
defined as the temperature at which $w_{1}(T_{p})=w_{-1}(T_{p})$.
To this end, one can use another Taylor series expansion of Boltzmann
factors around $\beta\rightarrow\beta_{p}$, where $\beta-\beta_{p}=\frac{T_{p}-T}{T_{p}T}=\frac{1}{T}(1-\frac{T}{T_{p}})=\frac{\tau}{T}$.
Thus, the Boltzmann's factor can be expanded using Taylor series as
a function of the inverse temperature $\beta$ around $\beta_{p}$,
as follows 
\begin{equation}
w_{n}(\beta)=w_{n}(\beta_{p})+\frac{\tau}{T_{p}}\frac{\partial w_{n}(\beta)}{\partial\beta}\Bigr|_{\beta=\beta_{p}}+\mathcal{O}\left(\tau^{2}\right).
\end{equation}
Introducing the notation $w_{n}(\beta_{p})=\tilde{w}_{n}$ and $a_{n}\tilde{w}_{n}=\frac{\partial w_{n}(\beta)}{T_{p}\,\partial\beta}\Bigr|_{\beta=\beta_{p}}$
the above equation can be simplified to 
\begin{alignat}{1}
w_{n}(\beta)= & \tilde{w}_{n}+a_{n}\,\tilde{w}_{n}\tau+\mathcal{O}\left(\tau^{2}\right),\nonumber \\
= & \tilde{w}_{n}\left(1+a_{n}\tau\right)+\mathcal{O}\left(\tau^{2}\right),
\end{alignat}
Further, let us express the expression $w_{1}-w_{-1}$ entering into
the denominator of Eq.~\eqref{f-apx} using this expansion 
\begin{equation}
w_{1}-w_{-1}=\tilde{w}_{1}\left[a_{1}-a_{-1}\right]\tau+\mathcal{O}\left(\tau{}^{2}\right).\label{eq:w1-w-1}
\end{equation}
From this formula one readily attains the following relation 
\begin{equation}
\left(a_{1}-a_{-1}\right)=\frac{1}{\tilde{w}_{1}T_{p}}\frac{\partial\left[w_{1}(\beta)-w_{-1}(\beta)\right]}{\partial\beta}\Bigr|_{\beta=\beta_{p}},\label{eq:coef-a1-a-1}
\end{equation}
which is quite helpful for obtaining the coefficients of power laws
pertinent to several physical quantities. An explicit formula for
this parameter is given by Eq.~ \eqref{eq:diff-a1s} in Appendix~\ref{appdxA}.
We emphasize that the development of power-law behavior is conditioned
to Eq.\eqref{cc} which implies that it is expected to hold when $\tau>\tilde{w}_{0}/(\tilde{w}_{1}|a_{1}-a_{-1}|)$.
The condition $\tau\rightarrow0$ implies that $\frac{\tilde{w}_{0}}{\tilde{w}_{1}|a_{1}-a_{-1}|}\rightarrow0$,
consequently, we must have $a_{1}\ne a_{-1}$. Therefore, it fails
very close to the quasicritical temperature at which the thermodynamic
functions are actually analytic.

\subsection{Correlation length}

The power-law behavior of the correlation length may be obtained analytically
by manipulating the relation \eqref{eq:coef-a1-a-1}. First, let us
rewrite $\frac{w_{1}}{w_{-1}}$ into the form 
\begin{alignat}{1}
\frac{w_{1}}{w_{-1}}= & 1+\left(a_{1}-a_{-1}\right)\tau+\mathcal{O}(\tau^{2}).\label{eq:w1/w-1}
\end{alignat}
Furthermore, one gets the following expression by performing the logarithm
of Eq. \eqref{eq:w1/w-1} in the limit of $\tau\rightarrow0$ 
\begin{alignat}{1}
\ln\left(\tfrac{w_{1}}{w_{-1}}\right)= & \ln\left[1+\left(a_{1}-a_{-1}\right)\tau\right]+\mathcal{O}(\tau^{2})\nonumber \\
= & \left(a_{1}-a_{-1}\right)\tau+\mathcal{O}(\tau^{2}).\label{eq:lnw1s}
\end{alignat}
The correlation length close to the quasitransition can be expressed
as follows 
\begin{equation}
\xi(\tau)=\left(\ln\tfrac{\lambda_{+}}{\lambda_{-}}\right)^{-1}=\begin{cases}
\left(\ln\tfrac{w_{1}}{w_{-1}}\right)^{-1}, & w_{1}>w_{-1}\\
\left(\ln\tfrac{w_{-1}}{w_{1}}\right)^{-1}, & w_{1}<w_{-1}
\end{cases}.\label{eq:Cr-gen}
\end{equation}
Using the leading-order term as given by Eq. \eqref{eq:lnw1s}, the
correlation length \eqref{eq:Cr-gen} reduces in general to 
\begin{equation}
\xi(\tau)=c_{_{\xi}}\;|\tau|^{-1}+\mathcal{O}(\tau^{0}),\label{eq:Cr-lght}
\end{equation}
where $c_{_{\xi}}=\frac{1}{|a_{1}-a_{-1}|}$ is constant independent
of temperature. Consequently, around the quasicritical temperature,
the correlation length generally follows the power-law function 
\begin{equation}
\xi(\tau)\propto|\tau|{}^{-\nu},
\end{equation}
whereas the relevant quasicritical exponent becomes $\nu=1$. We
recall that this result fails very near the quasicritical point at
which the correlation length remains finite. However, there may have
a finite range of temperatures in the close vicinity of the quasicritical
point on which a clear power-law behavior may develop, as we will
illustrate in the forthcoming sections.

\subsection{Specific heat}

Another physical quantity of interest is the specific heat and its
quasicritical exponents $\alpha$. To determine the quasicritical
behavior of the specific heat, let us at first rewrite the free energy
\eqref{eq:free-energ-2} for $w_{1}>w_{-1}$, and using the relation
\eqref{eq:w1-w-1} in the following form 
\begin{equation}
f=-T_{p}(1-\tau)\ln(1+a_{1}\tau)-\frac{T_{p}(1-\tau)\tilde{w}_{0}^{2}(1+a_{0}\tau)^{2}}{\tilde{w}_{1}^{2}(1+a_{1}\tau)(a_{1}-a_{-1})\tau}.
\end{equation}
By considering only the leading-order term from the Taylor series
expansion, the free energy reduces to 
\begin{alignat}{1}
f= & -\left(\frac{\tilde{w}_{0}}{\tilde{w}_{1}}\right)^{2}\frac{T_{p}}{(a_{1}-a_{-1})}\tau^{-1}+{\cal O}(\tau^{0})\\
\approx & -c_{_{f}}\,\tau^{-1},
\end{alignat}
where $c_{_{f}}=\left(\frac{\tilde{w}_{0}}{\tilde{w}_{1}}\right)^{2}\frac{T_{p}}{(a_{1}-a_{-1})}$
is a constant independent of temperature. For $w_{1}<w_{-1}$, we
have a very similar expression $c_{_{f}}=\left(\frac{\tilde{w}_{0}}{\tilde{w}_{1}}\right)^{2}\frac{T_{p}}{(a_{-1}-a_{1})}$.

Now, one may perform a derivative of the free energy with respect
to temperature. In doing so, one gets the following expression for
the entropy as a function of the temperature 
\begin{alignat}{1}
\mathcal{S}(\tau)= & -\left(\frac{\partial f}{\partial\tau}\right)\left(\frac{\partial\tau}{\partial T}\right)=-c_{_{f}}\,\tau^{-2}\left(\frac{-1}{T_{p}}\right)\nonumber \\
= & \frac{c_{_{f}}}{T_{p}}\,\tau^{-2}.
\end{alignat}
The above equation can be straightforwardly used in order to obtain
the formula governing temperature variations of the specific heat
in a vicinity of the quasicritical temperature 
\begin{equation}
C(\tau)=T\left(\frac{\partial\mathcal{S}}{\partial\tau}\right)\left(\frac{\partial\tau}{\partial T}\right)=2\frac{c_{_{f}}}{T_{p}}\,\tau^{-3}.\label{eq:Csp-exp}
\end{equation}
It is obvious from Eq.~\eqref{eq:Csp-exp} that, around the quasicritical
temperature, the specific heat follows the power law 
\begin{equation}
C(\tau)\propto|\tau|{}^{-\alpha},
\end{equation}
whereas the relevant quasicritical exponent is $\alpha=3$. Again,
this singularity becomes rounded as one ultimately approaches the
quasicritical temperature.

\subsection{Magnetic Susceptibility}

Last but not least, let us explore the power-law behavior of the magnetic
susceptibility around the quasicritical temperature. For this aim,
we will at first derive the explicit formula for the magnetization
\begin{alignat}{1}
M(\tau,h)=- & \left(\frac{\partial f}{\partial\tau}\right)\left(\frac{\partial\tau}{\partial h}\right)=-c_{_{f}}\,\left(\frac{\partial\tau}{\partial h}\Bigr|_{h_{p},T_{p}}\right)\tau^{-2}.\label{eq:Mng}
\end{alignat}
It is important to note that the parameters $T_{p}$ and $h_{p}$
are constrained by the relation $w_{1}(T_{p},h_{p})=w_{-1}(T_{p},h_{p})$,
which was denoted merely as $\tilde{w}_{1}=\tilde{w}_{-1}$. The isothermal
susceptibility is determined in the vicinity of the quasicritical
temperature just by the lowest-order term from the Taylor series expansion
\begin{alignat}{1}
\chi(\tau,h)= & \left(\frac{\partial M}{\partial\tau}\right)\frac{\partial\tau}{\partial h}=2c_{_{f}}\,\left(\frac{\partial\tau}{\partial h}\Bigr|_{h_{p},T_{p}}\right)^{2}\tau^{-3},\label{eq:Xi-tau}
\end{alignat}
where 
\begin{equation}
\frac{\partial\tau}{\partial h}\Bigr|_{h_{p},T_{p}}=\frac{w_{1,h_{p}}-w_{-1,h_{p}}}{\tilde{w}_{1}\left(a_{1}-a_{-1}\right)},\label{eq:tau-h}
\end{equation}
with $w_{1,h_{p}}=\frac{\partial w_{1}}{\partial h}|_{h_{p}}$ and
$w_{-1,h_{p}}=\frac{\partial w_{-1}}{\partial h}|_{h_{p}}$. The Eq.\eqref{eq:tau-h}
is valid for both condition $w_{1}>w_{-1}$ or $w_{-1}>w_{1}$. Accordingly,
the magnetic susceptibility follows the power law 
\begin{equation}
\chi(\tau)\propto|\tau|{}^{-\gamma},
\end{equation}
around the quasicritical temperature, whereas the relevant quasicritical
exponent is $\gamma=3$. This power-law behavior ultimately rounds
in the very close vicinity of the quasicritical temperature at which
the magnetic susceptibility remains finite. Notice that the quasicritical
temperature occurs for all physical observables at same point, the
quasicritical temperature can be obtain using the condition $w_{1}(T_{p})=w_{-1}(T_{p})$.

\section{Applications}

\begin{figure}
\includegraphics[scale=0.65]{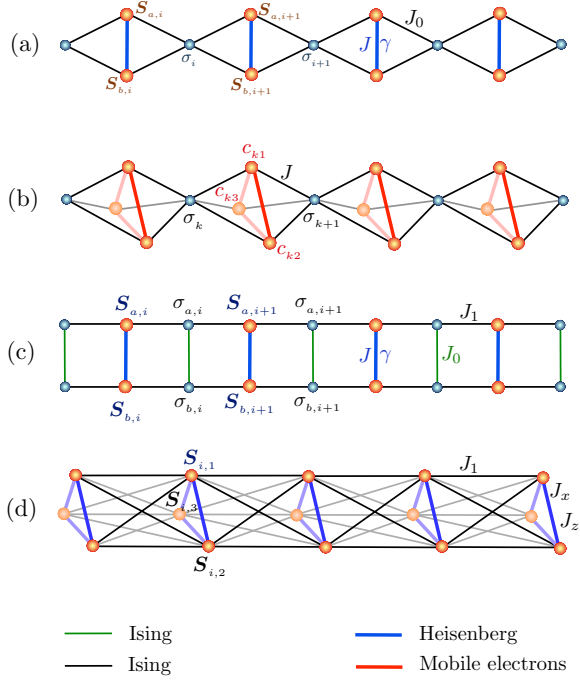} \caption{\label{fig:Models} A schematic illustration of four considered one-dimensional
models displaying a quasitransition: (a) Ising-XYZ diamond chain;
(b) coupled spin-electron double-tetrahedral chain; (c) Ising-Heisenberg
two-leg ladder; (d) Ising-Heisenberg three-leg tube. The former two
models belong to a class of one-dimensional models shown in Fig. \ref{fig0},
which can be rigorously mapped onto the effective Ising chain. The
latter two models do not belong to this class.}
\end{figure}

In this section, we will compare the quasicritical exponents as obtained
in the previous section from the approximate Taylor series expansion
performed around the quasicritical temperature with the relevant
exact results for two paradigmatic exactly solved models shown in
Fig.~\ref{fig:Models}(a)-(b), which can be rigorously mapped onto
the effective Ising chain. More specifically, we will comprehensively
explore the quasitransition of the spin-1/2 Ising-XYZ diamond chain
\cite{tor16} shown in Fig.~\ref{fig:Models}(a) and the coupled
spin-electron double-tetrahedral chain \cite{gal15} depicted in Fig.~\ref{fig:Models}(b),
respectively.

\subsection{Ising-XYZ diamond chain}

The spin-1/2 Ising-XYZ diamond chain has been introduced and exactly
solved in Ref. \cite{lis14}, whereas its quasitransition has been
discovered and detailed examined in Refs. \cite{tor16,sou17}. This
model schematically shown in Fig. \ref{fig:Models}(a) assumes a regular
alternation of the Ising spins $S_{i}=1/2$ with a couple of the Heisenberg
spins described by the Pauli spin operators $\sigma_{a(b),i}^{\alpha}$
($\alpha=\{x,y,z\}$), whereas the relevant Hamiltonian reads 
\begin{alignat}{1}
H= & -\sum_{i=1}^{N}\left[J(1+\gamma)\sigma_{a,i}^{x}\sigma_{b,i}^{x}+J(1-\gamma)\sigma_{a,i}^{y}\sigma_{b,i}^{y}+\right.\nonumber \\
 & +J_{z}\sigma_{a,i}^{z}\sigma_{b,i}^{z}+J_{0}(\sigma_{a,i}^{z}+\sigma_{b,i}^{z})(S_{i}+S_{i+1})+\nonumber \\
 & \left.+h_{0}(\sigma_{a,i}^{z}+\sigma_{b,i}^{z})+\frac{h}{2}(S_{i}+S_{i+1})\right].\label{eq:Hamt}
\end{alignat}
Above, the parameter $J_{0}$ denotes the Ising exchange interaction
between the nearest-neighbor Ising and Heisenberg spins, the XYZ exchange
coupling between the nearest-neighbor Heisenberg spin pairs is given
by three coupling constants: $J_{z}$ corresponding to the $z$-component,
$J$ corresponding to the $xy$-component and $\gamma$ being the
XY-anisotropy. Besides, the effect of external magnetic field $h$
($h_{0}$) acting on the Heisenberg spins (Ising spins) is considered
as well.

It turns out that the free energy \eqref{eq:free-energ-1} of this
model can be expressed in terms of the relevant Boltzmann factors,
which are given by the following relations (see Ref.~\cite{sou17}
for further details) 
\begin{equation}
w_{n}=2{\rm e}^{\frac{\beta nh_{0}}{2}}\left[{\rm e}^{-\frac{\beta J_{z}}{4}}{\rm cosh}\left(\tfrac{\beta J}{2}\right)+{\rm e}^{\frac{\beta J_{z}}{4}}{\rm cosh}\left(\beta\Delta_{n}\right)\right].
\end{equation}
with $\Delta_{n}=\sqrt{\left(h+J_{0}n\right)^{2}+\tfrac{1}{4}J^{2}\gamma^{2}}$
and $n=-1,0,1$.

\begin{figure}
\includegraphics[scale=0.21]{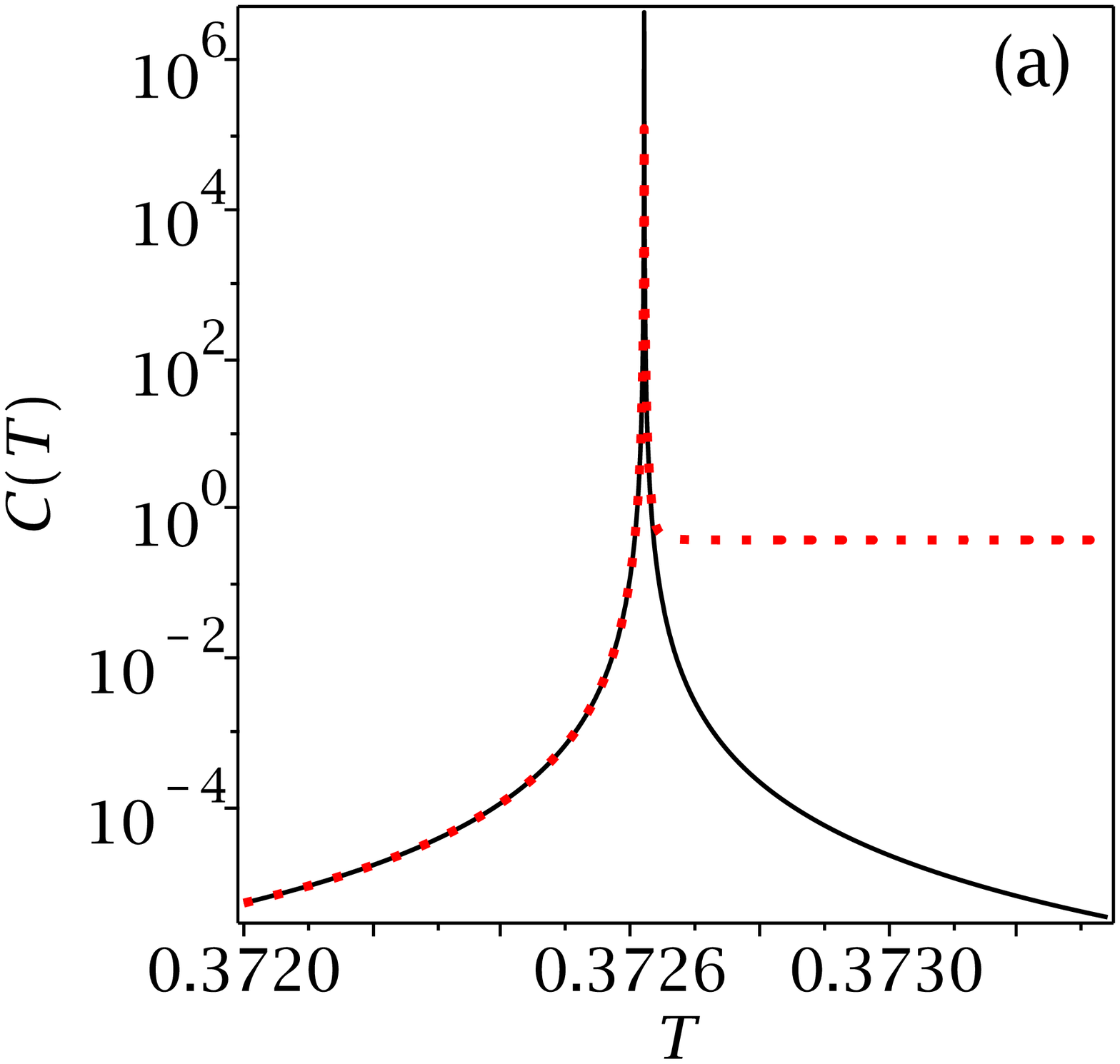}\includegraphics[scale=0.21]{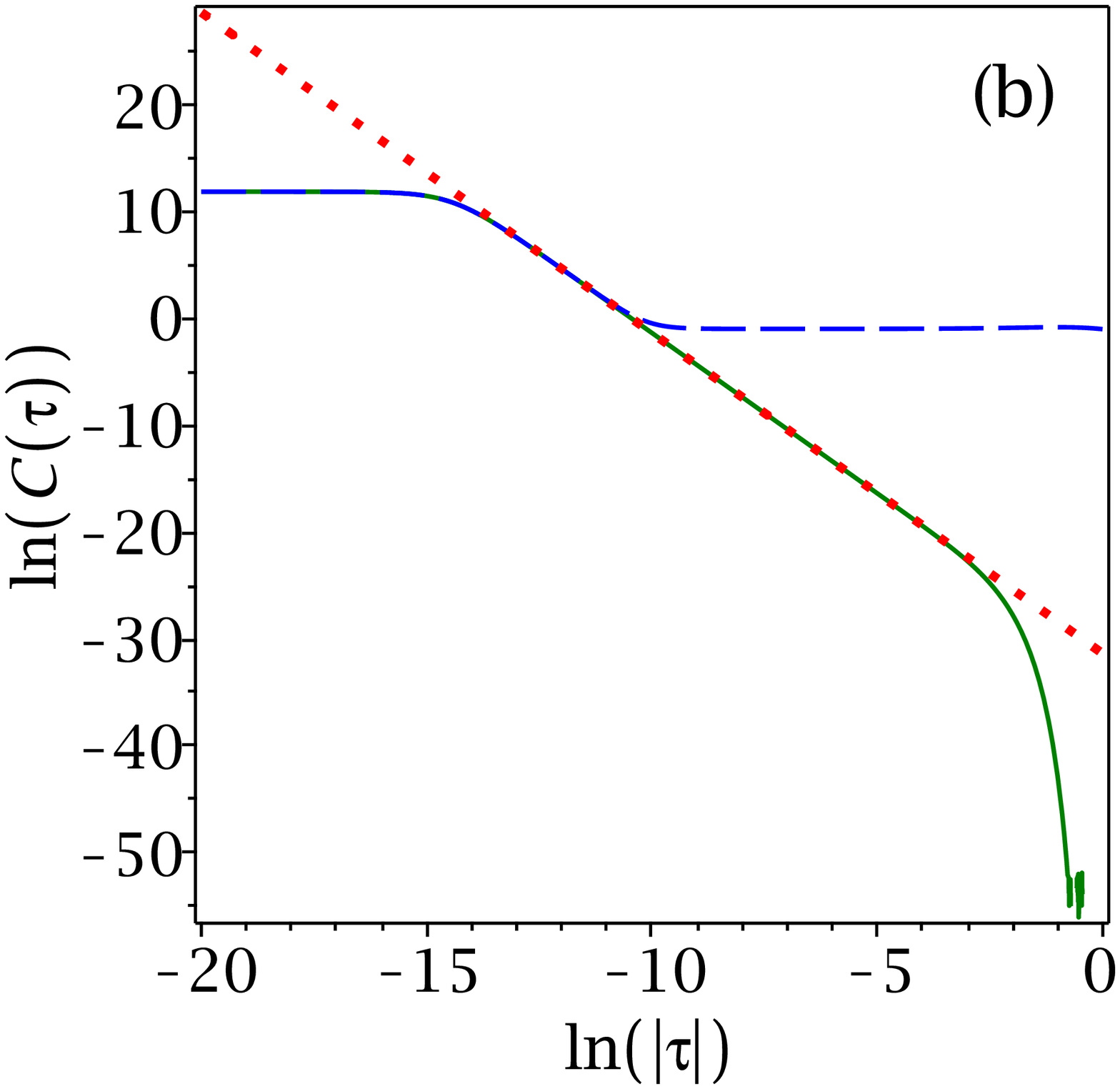}
\includegraphics[scale=0.21]{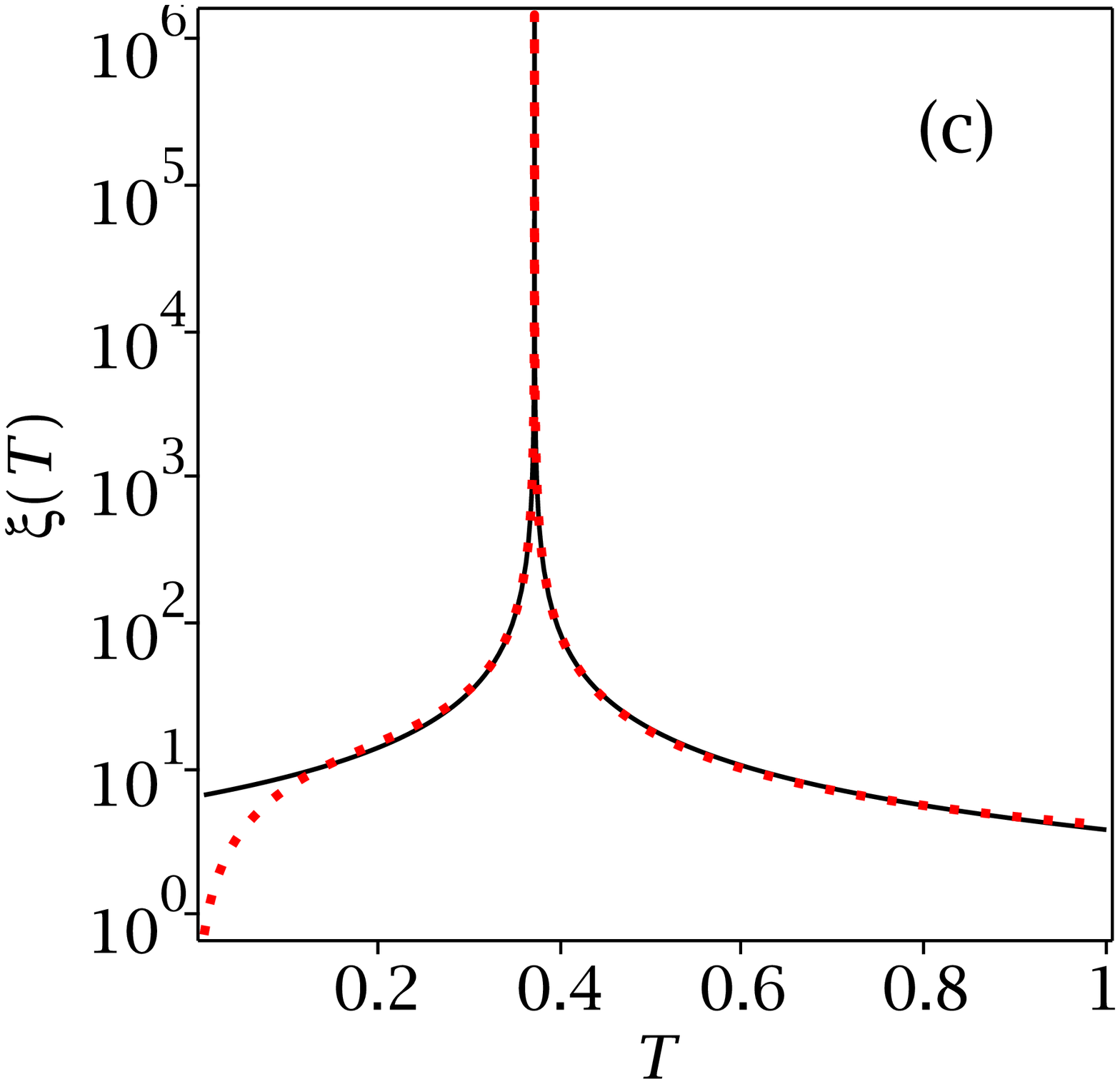}\includegraphics[scale=0.21]{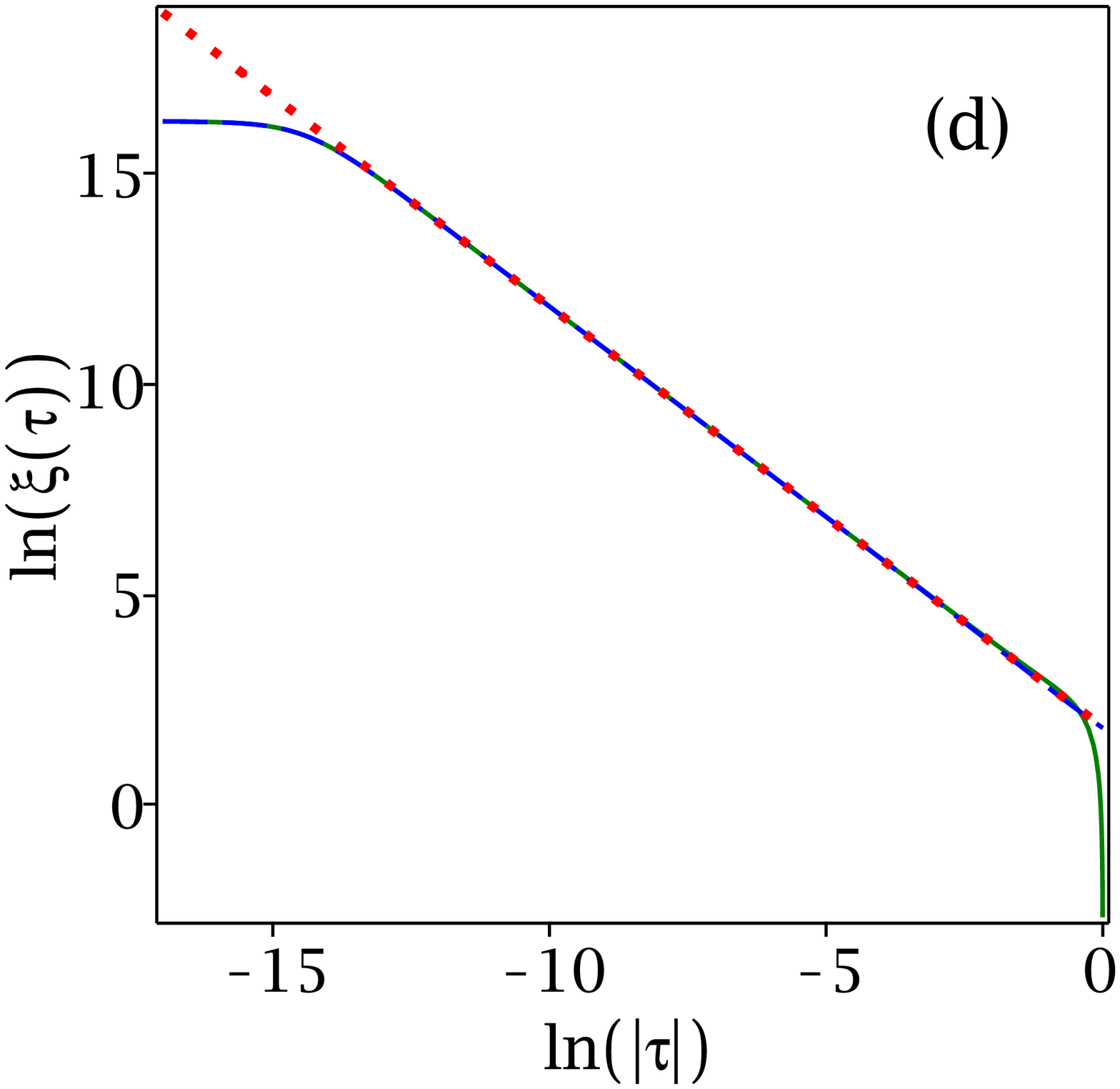}
\caption{\label{fig-xyz}Temperature variations of the specific heat and correlation
length of the spin-1/2 Ising-XYZ diamond chain (\ref{eq:Hamt}) in
a semi-logarithmic (left panel) and logarithmic (right panel) scale
in a vicinity of the quasicritical temperature for the particular
set of the interaction parameters: $J=100$, $J_{z}=24$, $J_{0}=-24$,
$\gamma=0.7$ and $h=12.7$. Left panel: solid lines correspond to
exact results, while dotted lines correspond to Taylor series expansion
as given by Eqs.~\eqref{eq:Csp-exp} and \eqref{eq:Cr-lght}, respectively.
Right panel: solid (dashed) lines correspond to exact results for
$\tau<0$ ($\tau>0$), while dotted lines are the power-law functions
$\ln(C(\tau))=-3\ln(|\tau|)-31.37$ and $\ln(\xi(\tau))=-\ln(|\tau|)+1.86$,
respectively.}
\end{figure}

Typical temperature variations of the specific heat and correlation
length of the spin-1/2 Ising-XYZ diamond chain are reported in Fig.~\ref{fig-xyz}
for the set of parameters $J=100$, $J_{z}=24$, $J_{0}=-24$, $\gamma=0.7$,
and $h=12.7$, which are consistent with emergence of a quasitransition
at the quasicritical temperature $T_{p}=0.37262119$. The readers
interested in further details concerning the specific heat and correlation
length are referred to Ref. \cite{sou17}. In Fig.~\ref{fig-xyz}(a)
the specific heat $C(T)$ is plotted against temperature $T$, whereas
a solid line represents exact results as given in Refs. \cite{tor16,sou17}
and a dotted line denotes Taylor series expansion around the quasicritical
temperature as given by Eq.~\eqref{eq:Csp-exp}. The temperature
dependence of $\ln(C(\tau))$ versus $\ln(|\tau|)$ depicted in Fig.~\ref{fig-xyz}(b)
verifies existence of intermediate temperature range, where the specific
heat follows the power law (a straight line in log-log scale) with
the critical exponent $\alpha=\alpha'=3$. Exact results for the specific
heat are indeed consistent with $\ln(C(\tau))=-3\ln(|\tau|)-31.37$
as obtained from Taylor series expansion given by Eq.\eqref{eq:Csp-exp}.
Furthermore, the correlation length $\xi(T)$ is displayed against
$T$ in Fig.~\ref{fig-xyz}(c), where the relevant exact results
are depicted by a solid line and the Taylor series expansion given
by Eq.~\eqref{eq:Cr-lght} by a dotted line. It can be seen from
$\ln(\xi(\tau))$ vs. $\ln(|\tau|)$ dependence shown in Fig.~\ref{fig-xyz}(d)
that the correlation length follows sufficiently close but not too
close to a quasicritical temperature the power law with the critical
exponent $\nu=\nu'=1$. In fact, the exact results reported in Refs.
\cite{tor16,sou17} are in reasonable accordance with the Taylor series
expansion as given by Eq.~\eqref{eq:Cr-lght} illustrated as a straight
dotted line.

\begin{figure}
\includegraphics[scale=0.21]{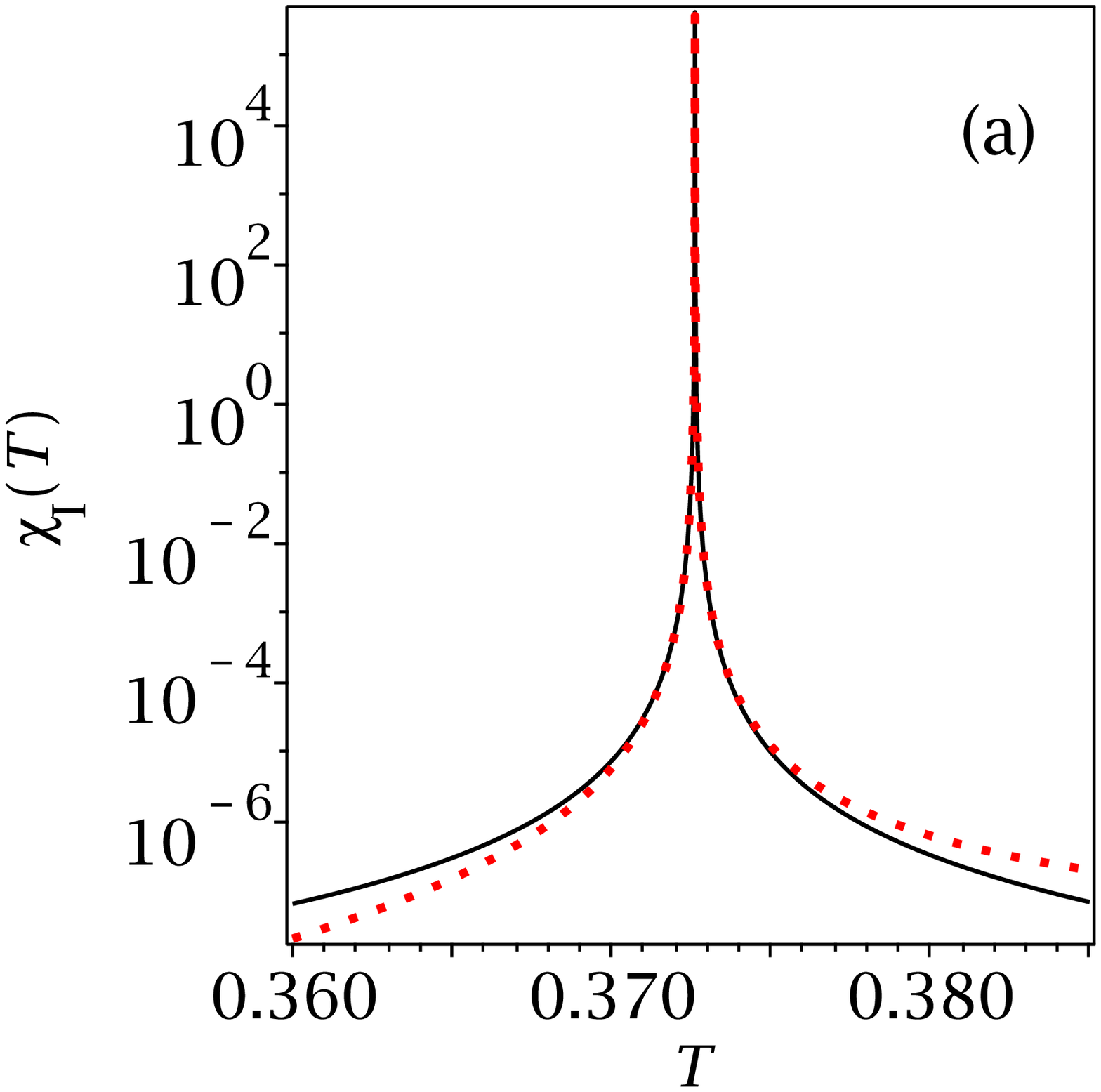}\includegraphics[scale=0.21]{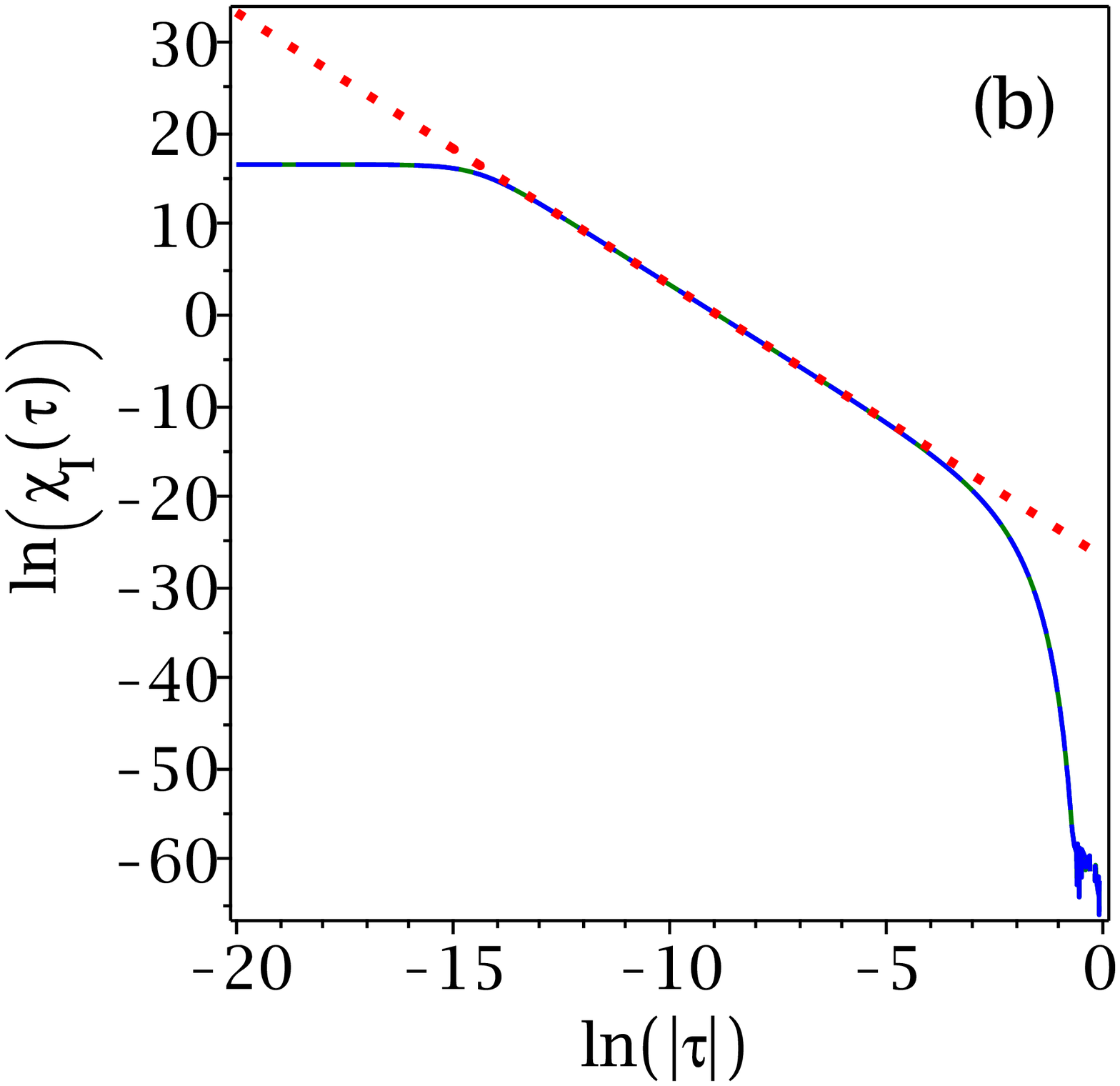}
\includegraphics[scale=0.21]{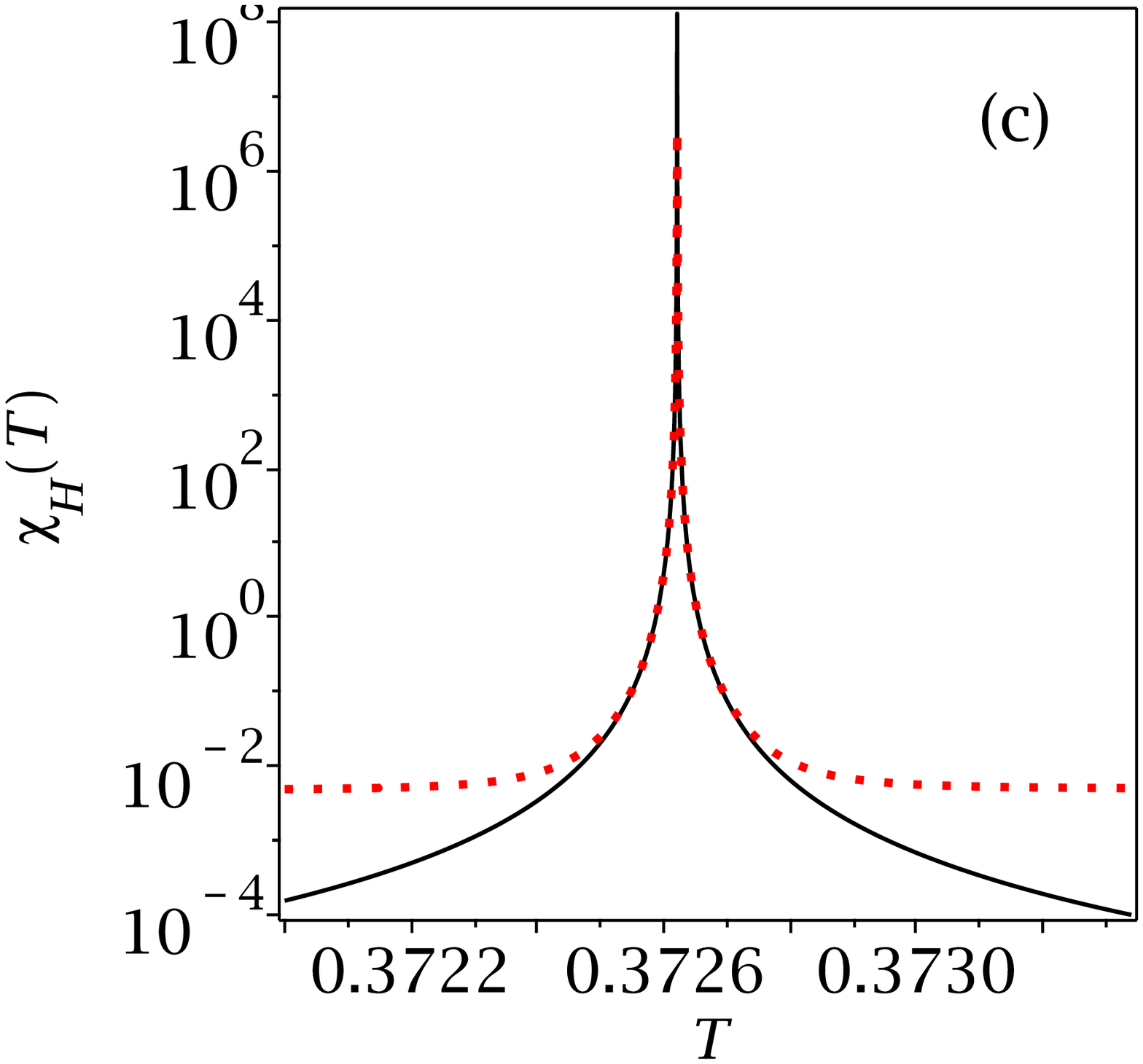}\includegraphics[scale=0.21]{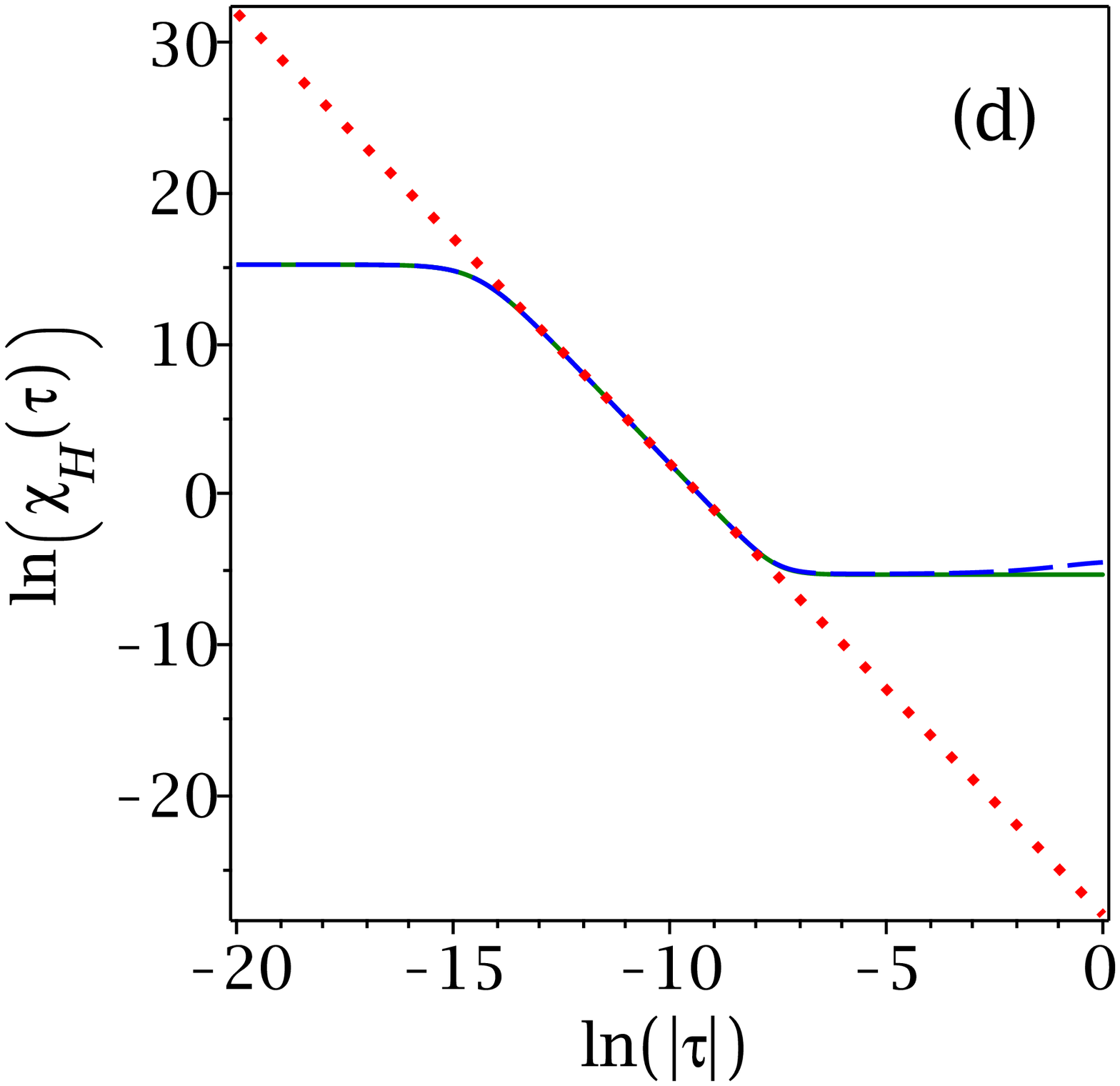}
\caption{\label{fig-xyz-Suscp}Temperature variations of the magnetic susceptibility
of the Ising ($\chi_{I}$) and Heisenberg ($\chi_{H}$) spins for
the spin-1/2 Ising-XYZ diamond chain (\ref{eq:Hamt}). The same parameter
set is used as in Fig.\ref{fig-xyz}. Left panel: solid lines correspond
to exact results, while dotted lines correspond to Taylor series expansion
as given by Eqs.~\eqref{eq:Xi-tau}. Right panel: solid (dashed)
lines correspond to exact results for $\tau<0$ ($\tau>0$), while
dotted lines are the power-law functions $\ln(\chi_{_{I}}(\tau))=-3\ln(|\tau|)-26.667$
and $\ln(\chi_{_{H}}(\tau))=-3\ln(|\tau|)-27.976$.}
\end{figure}

Last but not least, the magnetic susceptibility $\chi_{_{I}}(T)$
of the Ising spins is displayed in Fig.~\ref{fig-xyz-Suscp}(a) as
a function of temperature $T$, whereas a solid line refers to exact
results derived in Refs. \cite{tor16,sou17} and a dotted line labels
asymptotic expression as obtained from Taylor series expansion given
by Eq.~\eqref{eq:Xi-tau}. The magnetic susceptibility shown in Fig.~\ref{fig-xyz-Suscp}(b)
in the form $\ln(\chi_{_{I}}(\tau))$ against $\ln(|\tau|)$ plot
corroborates an intermediate temperature range, where the magnetic
susceptibility follows the power law with the critical exponent $\gamma=\gamma'=3$.
The dotted line, which was obtained from Taylor series expansion given
by Eq.~\eqref{eq:Xi-tau} with the asymptotic form $\ln(\chi_{_{I}}(\tau))=-3\ln(|\tau|)-26.667$,
is in this temperature range in a plausible agreement with the exact
results. Similar findings hold for the magnetic susceptibility of
the Heisenberg spin $\chi_{_{H}}(T)$, which is illustrated in Fig.\ref{fig-xyz-Suscp}(c)
and \ref{fig-xyz-Suscp}(d). However, it is worth noticing that the
magnetic susceptibility of the Ising spins follows the relevant power-law
function in a wider temperature region as compared to the magnetic
susceptibility of the heisenberg spins.

It could be concluded that the specific heat, magnetic susceptibility
and correlation length of the spin-1/2 Ising-XYZ diamond chain driven
sufficiently close but not too close to a quasicritical temperature
are characterized by the power-law functions with the critical exponents
$\alpha=\alpha'=3$, $\gamma=\gamma'=3$ and $\nu=\nu'=1$. Of course,
this description inevitably breaks down at the quasicritical temperature
as the system does not exhibit actual divergence of the relevant physical
quantities.

\subsection{Spin-electron double-tetrahedral chain}

Next, let us consider a coupled spin-electron model on a double-tetrahedral
chain schematically depicted in Fig. \ref{fig:Models}(b), in which
one localized Ising spin situated at nodal site regularly alternates
with a triangular plaquette composed of three decorating sites available
to two mobile electrons. This one-dimensional spin-electron system
has been introduced and exactly solved in Ref. \cite{gal15}, where
the outstanding temperature dependencies of several physical quantities
mimicking a phase transition were also reported. The coupled spin-electron
model on a double-tetrahedral chain can be defined as a sum over block
Hamiltonians ${\cal H}_{k}$ 
\begin{equation}
{\cal H}=\sum_{k=1}^{N}{\cal H}_{k},\label{eq:H}
\end{equation}
whereas each block Hamiltonian ${\cal H}_{k}$ involves all the interaction
terms connected to two mobile electrons delocalized over the $k$th
triangular plaquette 
\begin{eqnarray}
{\cal H}_{k}\!\! & = & \!\!-t\sum_{\alpha=\uparrow,\downarrow}\!(c_{k1,\alpha}^{\dagger}c_{k2,\alpha}\!+c_{k2,\alpha}^{\dagger}c_{k3,\alpha}\!+c_{k3,\alpha}^{\dagger}c_{k1,\alpha}\!+{\rm h.c.})\nonumber \\
\!\! & + & \!\!\frac{J}{2}(\sigma_{k}^{z}+\sigma_{k+1}^{z})\sum_{j=1}^{3}\,(n_{kj,\uparrow}-n_{kj,\downarrow})+U\sum_{j=1}^{3}n_{kj,\uparrow}n_{kj,\downarrow}\nonumber \\
\!\! & - & \!\!\frac{h_{{\rm I}}}{2}(\sigma_{k}^{z}+\sigma_{k+1}^{z})-\frac{h_{{\rm e}}}{2}\sum_{j=1}^{3}\,(n_{kj,\uparrow}-n_{kj,\downarrow}).\label{eq:Hk}
\end{eqnarray}
Here, $c_{kj,\alpha}^{\dagger}$ and $c_{kj,\alpha}$ label standard
fermionic creation and annihilation operators for mobile electrons
from the $k$th triangular plaquette with spin $\alpha$ = $\uparrow$
or $\downarrow$, $n_{kj,\alpha}=c_{kj,\alpha}^{\dagger}c_{kj,\alpha}$
is the respective number operator and $\sigma_{k}^{z}=\pm1/2$ denotes
the Ising spin situated at the $k$th nodal site. The hopping term
$t>0$ accounts for the kinetic energy of mobile electrons delocalized
over triangular plaquettes, the Coulomb term $U>0$ is energy penalty
for two electrons with opposite spins situated at the same decorating
site and the coupling constant $J$ determines the Ising-type nearest-neighbor
interaction between the localized Ising spins and the mobile electrons.
Finally, the Zeeman's terms $h_{{\rm I}}$ and $h_{{\rm e}}$ account
the magnetostatic energy of the localized Ising spins and mobile electrons
in a static magnetic field.

A diagonalization of the block Hamiltonian (\ref{eq:Hk}) gives a
full energy spectrum (see Eq. (5) in Ref. \cite{gal15}), whereas
the resulting expression for the relevant Boltzmann factor obtained
from this complete set of eigenvalues reads 
\begin{alignat}{1}
w_{n}= & {\rm e}^{\beta\frac{n}{2}h_{{\rm I}}}\Big\{\!\!\left(\!2{\rm e}^{\beta t}+{\rm e}^{-2\beta t}\right)\!\left(1+2\cosh\left[\beta(Jn-h_{{\rm e}})\right]\right)\nonumber \\
 & +\,4{\rm e}^{-\beta t/2-\beta U/2}\cosh\left[\frac{\beta}{2}\!\sqrt{(U\!-t)^{2}\!+8t^{2}}\,\right]\nonumber \\
 & +\negmedspace2{\rm e}^{\beta t-\beta U/2}\cosh\negmedspace\left[\frac{\beta}{2}\!\sqrt{(U\!+2t)^{2}\!+32t^{2}}\,\right]\negmedspace\Big\}
\end{alignat}
with the parameter $n=(\sigma_{k}^{z}+\sigma_{k+1}^{z})$ defined
for the sake of brevity. The free energy for the coupled spin-electron
double-tetrahedral chain can be consequently obtained from Eq. \eqref{eq:free-energ-1}
by assuming $w_{-1}$, $w_{0}$ and $w_{1}$.

\begin{figure}
\includegraphics[scale=0.21]{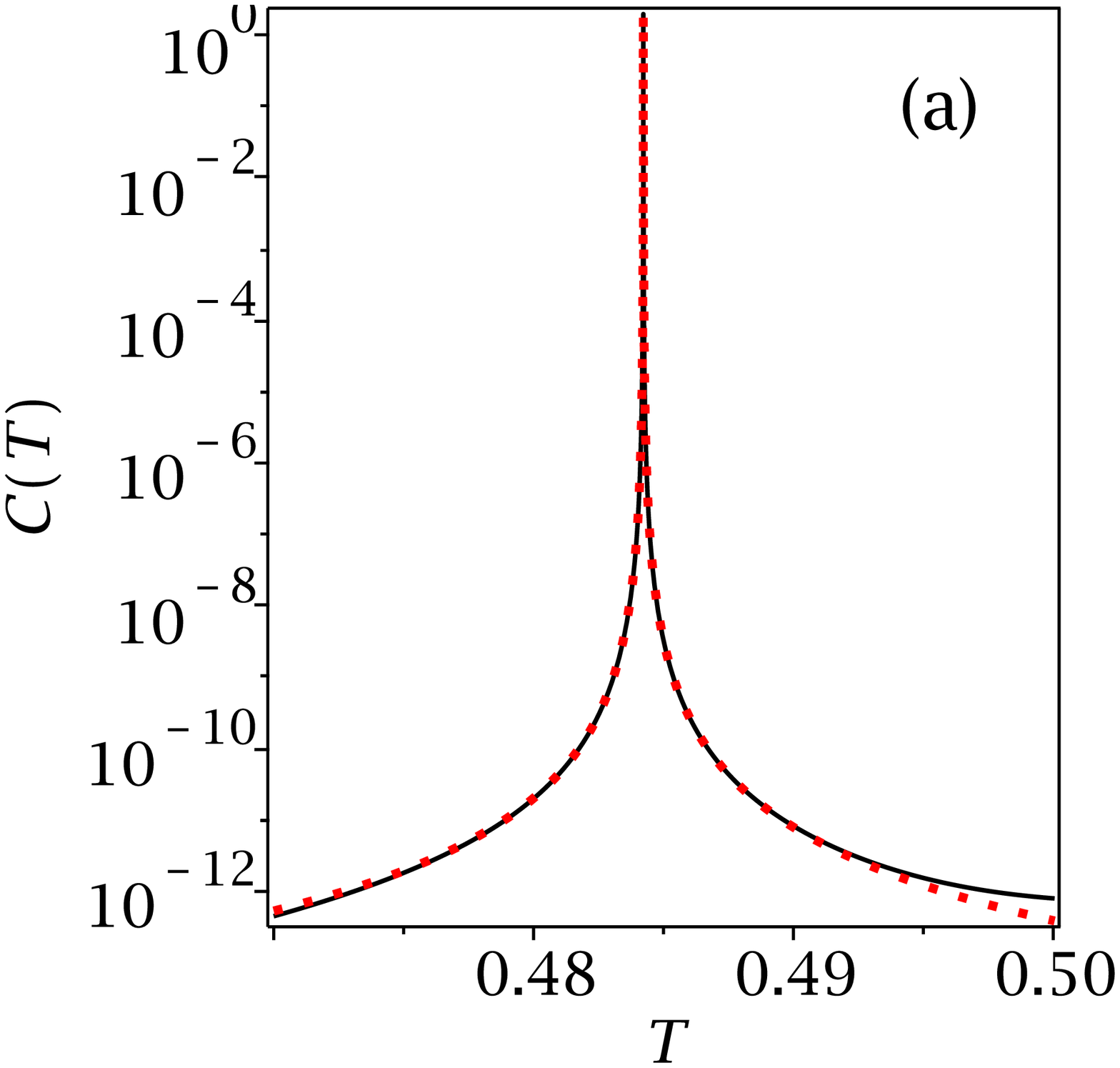}\includegraphics[scale=0.21]{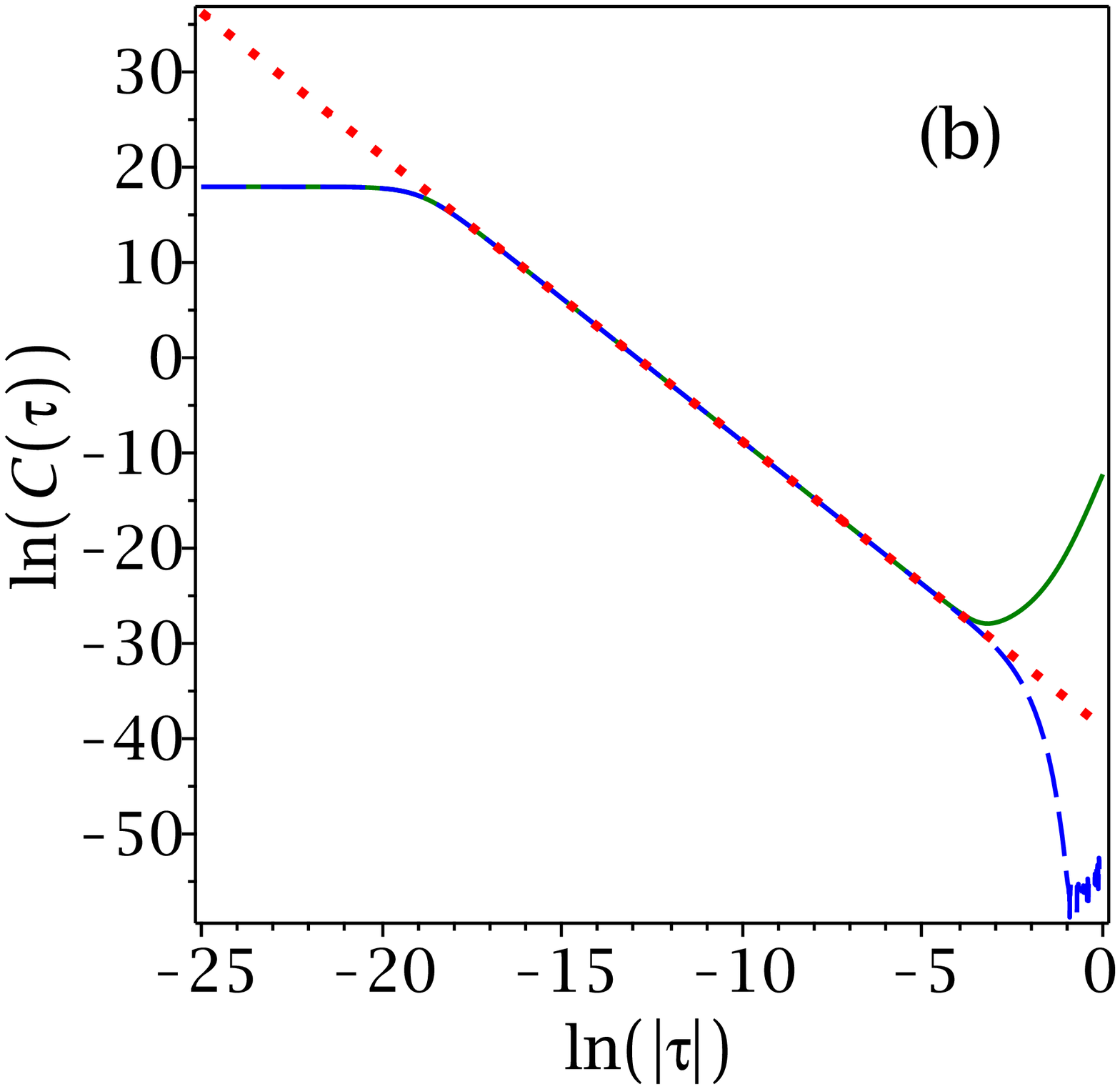}
\includegraphics[scale=0.21]{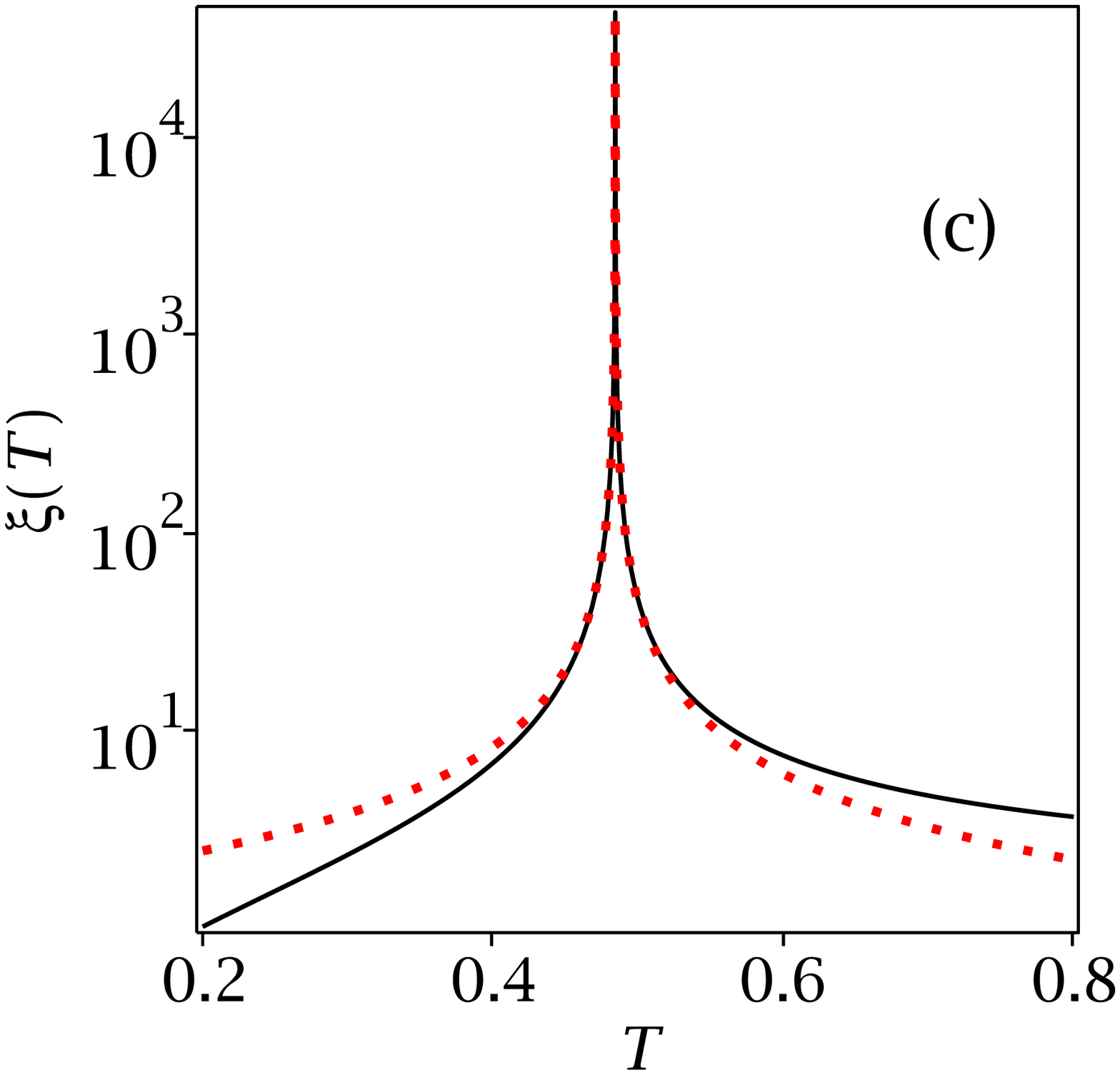}\includegraphics[scale=0.21]{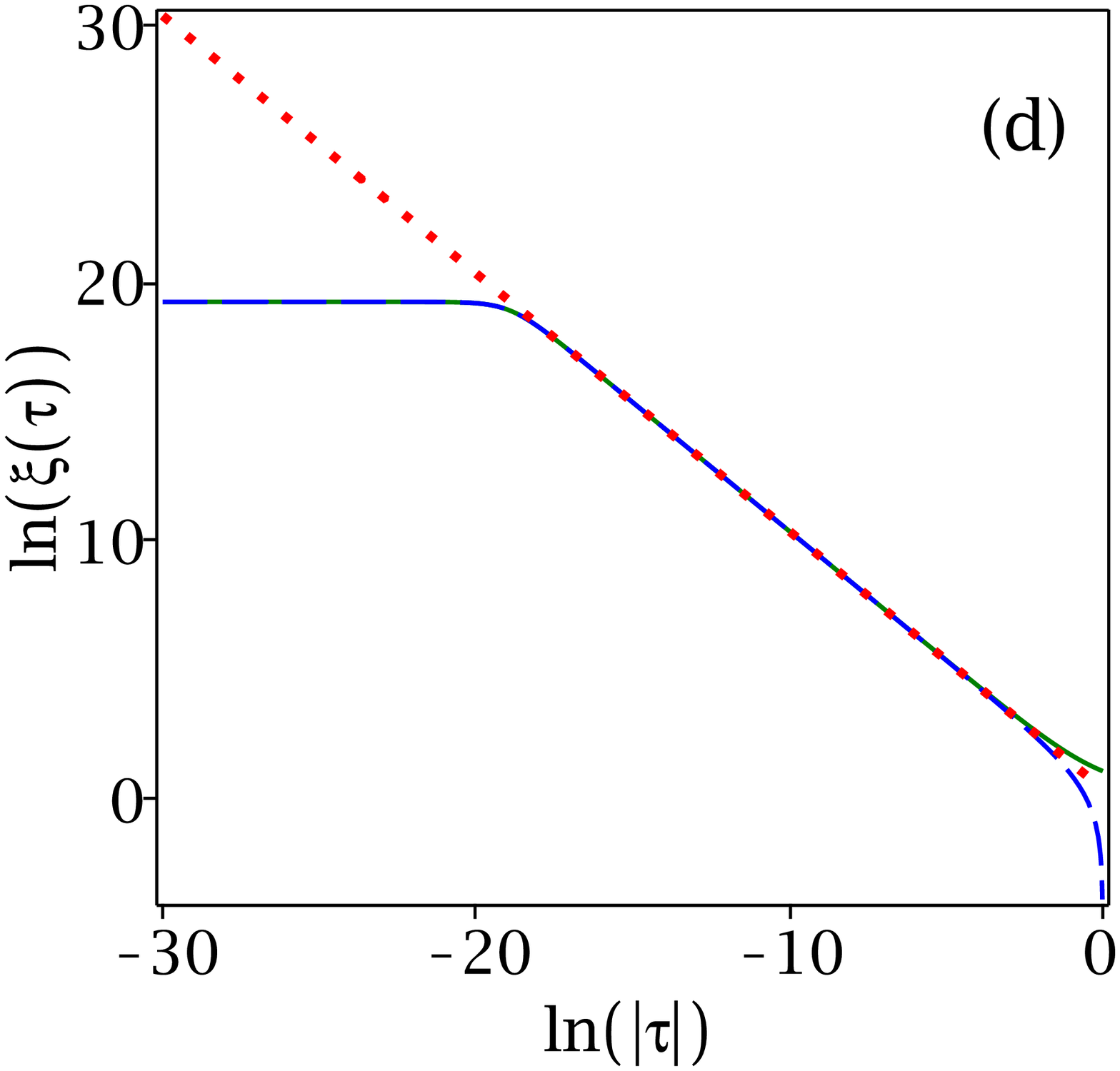}
\caption{\label{tetrahedral}Temperature variations of the specific heat and
correlation length of the coupled spin-electron double-tetrahedral
chain (\ref{eq:Hk}) in a semi-logarithmic (left panel) and logarithmic
(right panel) scale in a close vicinity of the quasitransition for
the particular set of the interaction parameters $t=8.5$, $U=20$
and $J=20$. Left panel: solid lines correspond to exact results,
while dotted lines correspond to Taylor series expansion as given
by Eqs.~\eqref{eq:Csp-exp} and \eqref{eq:Cr-lght}, respectively.
Right panel: solid (dashed) lines correspond to exact results for
$\tau<0$ ($\tau>0$), while dotted lines are the power-law functions
$\ln(C(\tau))=-3\ln(|\tau|)-38.86$ and $\ln(\xi(\tau))=-\ln(|\tau|)+0.366$,
respectively.}
\end{figure}

It has been argued in Ref. \cite{gal15} that the coupled spin-electron
double-tetrahedral chain given by the Hamiltonian (\ref{eq:Hk}) mimics
a phase transition at the quasicritical temperature 
\begin{eqnarray}
T_{p}=\frac{\sqrt{(U+2t)^{2}+32t^{2}}-U-2J}{\ln4}.\label{eq:ptc}
\end{eqnarray}
To illustrate the case, we depict in Fig.~\ref{tetrahedral} typical
temperature variations of the specific heat and correlation length
by assuming the set of interaction parameters $t=8.5$, $U=20$, $J=20$
and $h_{e}=h_{I}=20$, which lead to a quasitransition at the quasicritical
temperature $T_{p}=(\sqrt{3681}-60)/(\ln4)\approx0.4842011$. Fig.~\ref{tetrahedral}(a)
compares exact results for temperature dependence of the specific
heat $C(T)$ (solid line) derived according to Ref. \cite{gal15}
with the asymptotic formula \eqref{eq:Csp-exp} derived from Taylor
series expansion around the quasicritical temperature (dotted line).
It turns out that the specific heat actually follows sufficiently
close but not too close to the quasicritical temperature the power
law with the critical exponent $\alpha=\alpha'=3$ as it is evidenced
by a straight dotted line $\ln(C(\tau))=-3\ln(|\tau|)-38.8592$ shown
in Fig.~\ref{tetrahedral}(b) in the respective $\ln(C(\tau))$ vs.
$\ln(|\tau|)$ dependence. Similarly, exact results for temperature
dependence of the correlation length $\xi(T)$ (solid line) are plotted
in Fig.~\ref{tetrahedral}(c) along with asymptotic expression \eqref{eq:Cr-lght}
(dotted line) derived from the Taylor series expansion around the
quasicritical temperature. It is quite evident from $\ln(\xi(\tau))$
vs. $\ln(|\tau|)$ dependence shown in Fig.~\ref{tetrahedral}(d)
that the correlation length is governed the power law \eqref{eq:Cr-lght}
with the quasicritical exponent $\nu=\nu'=1$ if temperature is set
sufficiently close but not too close to a quasicritical one.

\begin{figure}
\includegraphics[scale=0.21]{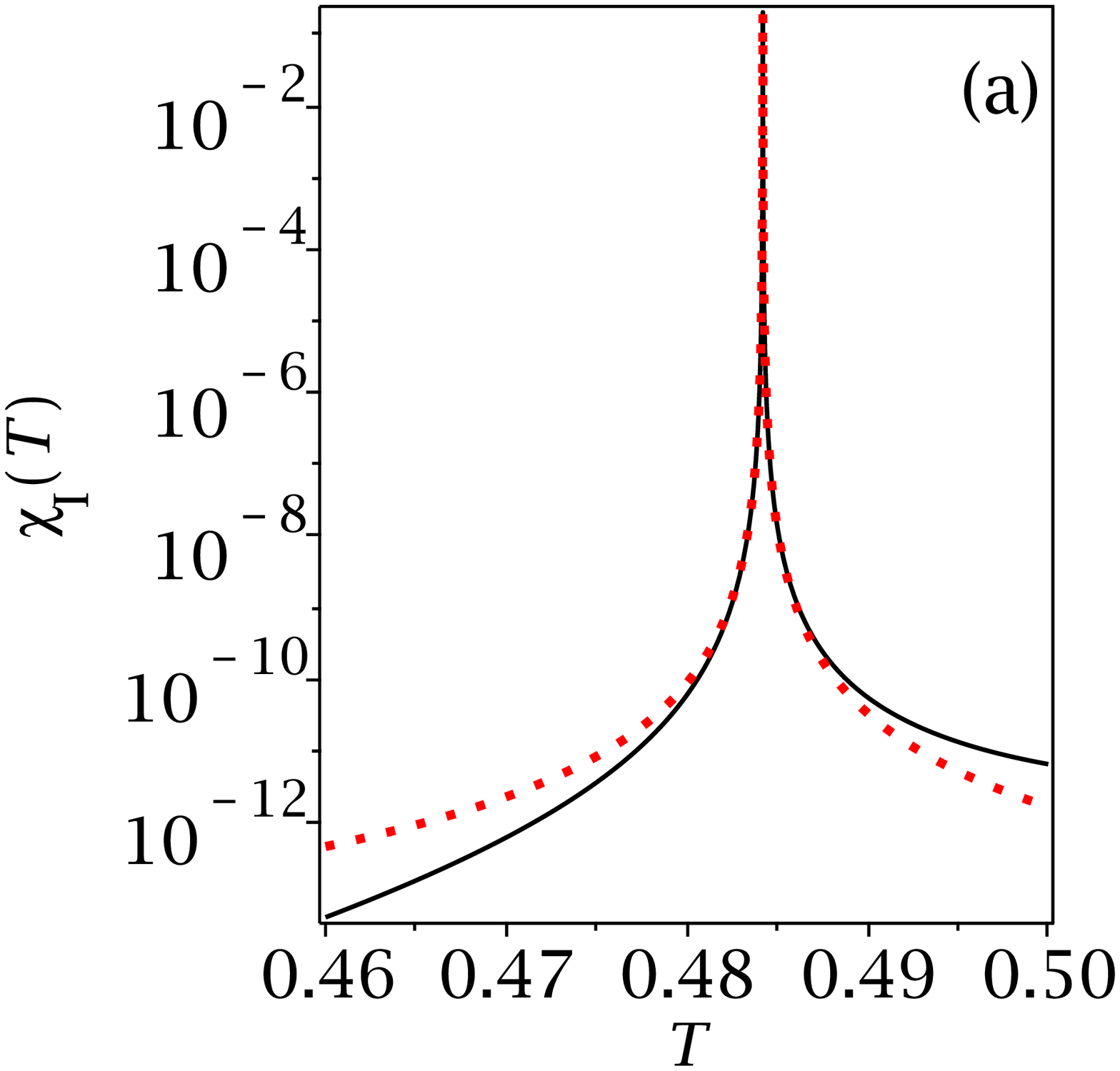}\includegraphics[scale=0.21]{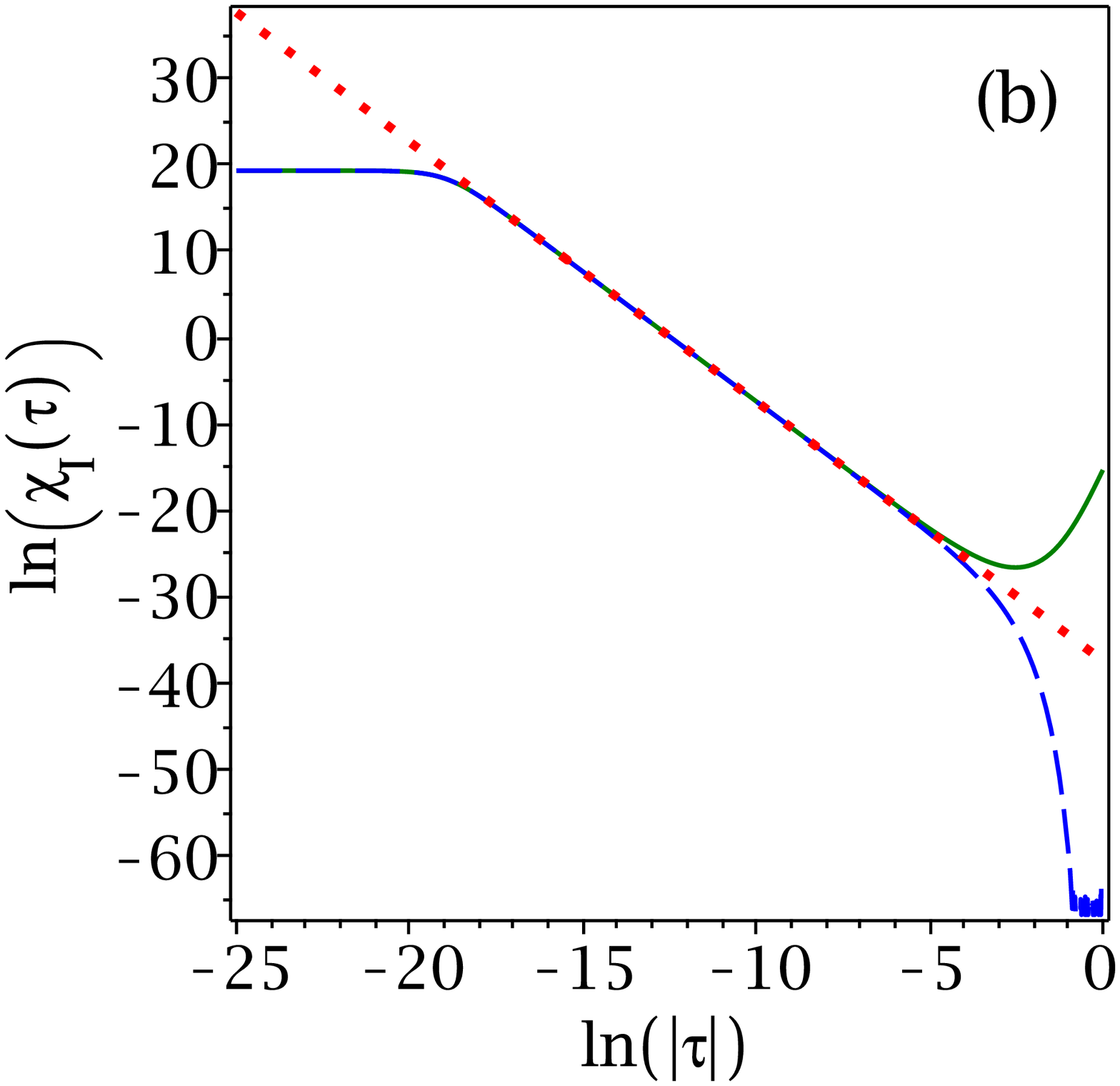}
\includegraphics[scale=0.21]{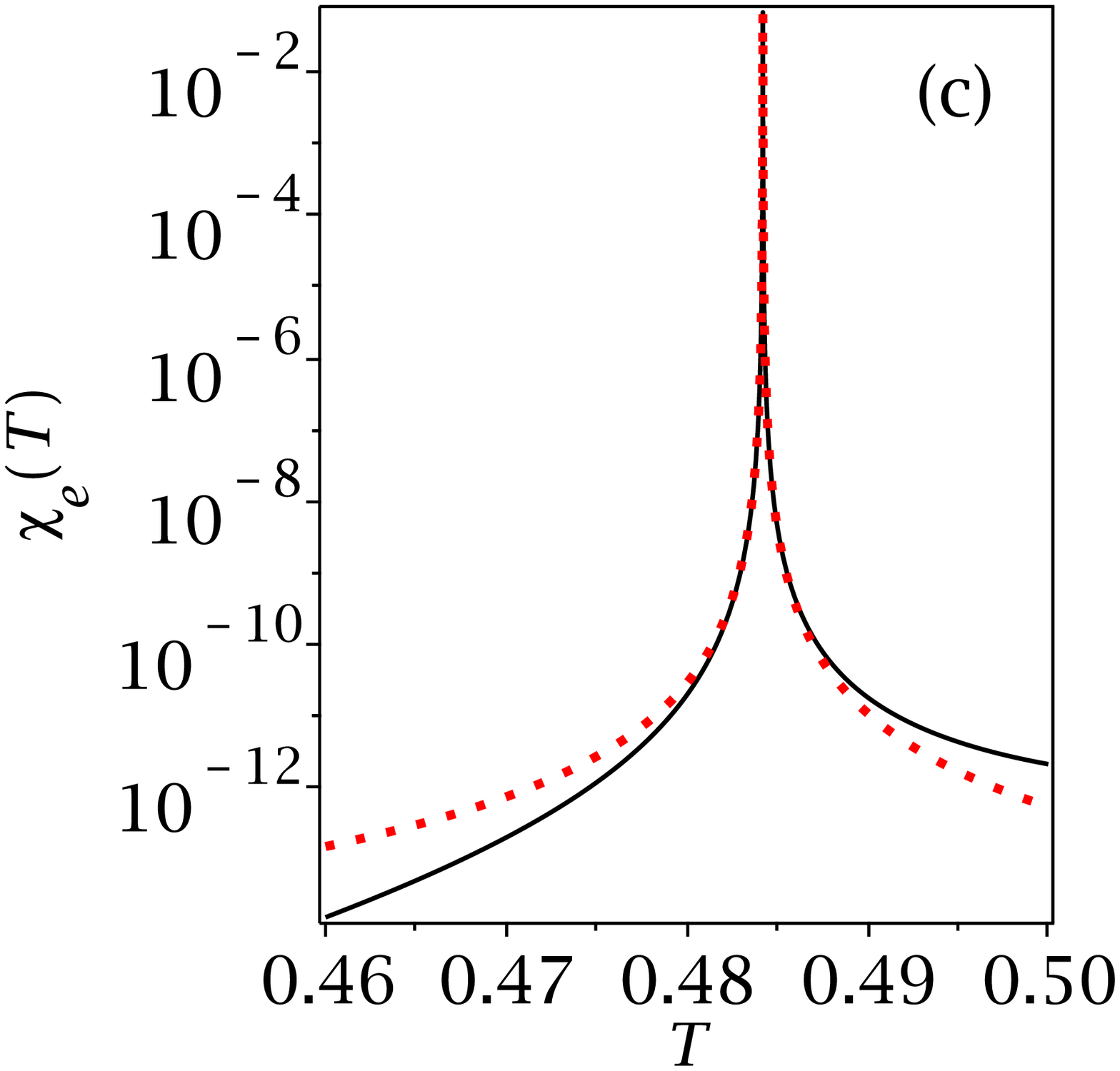}\includegraphics[scale=0.21]{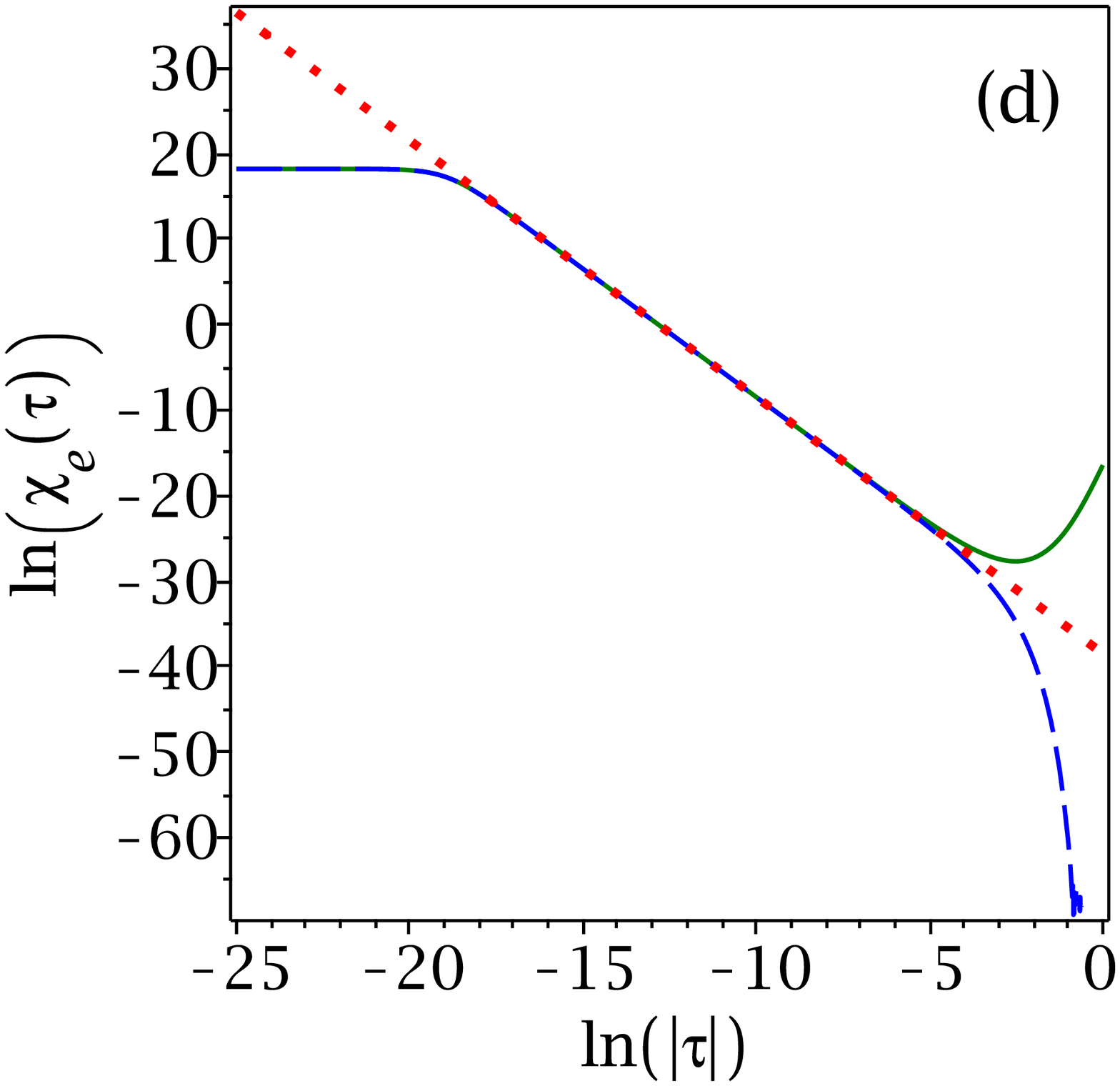}
\caption{\label{tetrahedral-Scpt}Temperature variations of the magnetic susceptibility
of the Ising spins ($\chi_{I}$) and the mobile electrons ($\chi_{e}$)
for the coupled spin-electron double-tetrahedral chain (\ref{eq:Hk}).
The same parameter set is used as in Fig.\ref{tetrahedral}. Left
panel: solid lines correspond to exact results, while dotted lines
correspond to Taylor series expansion as given by Eqs.~\eqref{eq:Xi-tau}.
Right panel: solid (dashed) lines correspond to exact results for
$\tau<0$ ($\tau>0$), while dotted lines are the power-law functions
$\ln(\chi_{_{I}}(\tau))=-3\ln(|\tau|)-38.40$ and $\ln(\chi_{e}(\tau))=-3\ln(|\tau|)-38.50$. }
\end{figure}

Finally, exact results (solid line) for temperature variations of
the magnetic susceptibility of the Ising spins $\chi_{_{I}}(T)$ depicted
in Fig.~\ref{tetrahedral-Scpt}(a) are in plausible concordance with
the asymptotic expression \eqref{eq:Xi-tau} obtained from Taylor
series expansion around the quasicritical temperature. In addition,
$\ln(\chi_{_{I}}(\tau))$ against $\ln(|\tau|)$ plot displayed in
Fig.~\ref{tetrahedral-Scpt}(b) verifies existence of an intermediate
temperature region, where exact results for the susceptibility (solid
line) follow the power law $\ln(\chi_{_{I}}(\tau))=-3\ln(|\tau|)-38.4009$
with the critical exponent $\gamma=\gamma'=3$ (dotted line) obtained
from Taylor series expansion \eqref{eq:Xi-tau}. Analogously, the
magnetic susceptibility of the mobile electrons is shown in Fig.\ref{tetrahedral-Scpt}(c)
and \ref{tetrahedral-Scpt}(d), where a similar coincidence is found
with the power-law dependence characterized through almost the same
constants $\ln(\chi_{e}(\tau))=-3\ln(|\tau|)-38.4995$.

To summarize, it has been found that the specific heat, magnetic susceptibility
and correlation length of the coupled spin-electron double-tetrahedral
chain are governed in a close vicinity of the quasicritical temperature
by the power-law functions, which are characterized by the same set
of quasicritical exponents $\alpha=\alpha'=3$, $\gamma=\gamma'=3$
and $\nu=\nu'=1$ as reported previously for the spin-1/2 Ising-XYZ
diamond chain even though both one-dimensional lattice-statistical
models are very different in their nature.

\section{''quasicriticality'' of other one-dimensional models}

In this section, we will comprehensively explore the quasicritical
exponents of other one-dimensional lattice-statistical models, which
cannot be in principle mapped onto the effective Ising chain. It will
be demonstrated hereafter that the quasicritical exponents of other
paradigmatic examples of one-dimensional models displaying a quasitransition
at finite temperatures will remain the same, which indicates a certain
universality of the quasitransitions. More specifically, we will
exactly validate quasicritical exponents of the spin-1/2 Ising-XXZ
two-leg ladder \cite{roj16} and the spin-1/2 Ising-XXZ three-leg
tube \cite{str16}, respectively.

\subsection{Ising-XXZ two-leg ladder}

First, let us examine quasicritical exponents of the the spin-1/2
Ising-XXZ two-leg ladder with regularly alternating Ising and Heisenberg
rungs as schematically represented in Fig.~\ref{fig:Models}(c).
The Hamiltonian of the investigated one-dimensional spin system can
be expressed by 
\begin{equation}
\mathcal{H}=\sum_{i=1}^{N}\left(H_{i}^{XXZ}+H_{i,i+1}^{I}+H_{i,i+1}^{IH}\right)\label{eq:Ham-orig}
\end{equation}
with 
\begin{alignat}{1}
H_{i}^{XXZ}= & -J_{x}(S_{a,i}^{x}S_{b,i}^{x}+S_{a,i}^{y}S_{b,i}^{y})-J_{z}S_{a,i}^{z}S_{b,i}^{z},\nonumber \\
H_{i,i+1}^{I}= & -\frac{J_{0}}{2}(\sigma_{a,i}\sigma_{b,i}+\sigma_{a,i+1}\sigma_{b,i+1}),\nonumber \\
H_{i,i+1}^{IH}= & -J_{1}(\sigma_{a,i}+\sigma_{a,i+1})S_{a,i}^{z}-J_{1}(\sigma_{b,i}+\sigma_{b,i+1})S_{b,i}^{z}.\label{eq:Ht-HI}
\end{alignat}
Here, $S_{\gamma,i}^{\alpha}$ ($\alpha=\{x,y,z\}$) denote three
spatial components of the spin-1/2 operator pertinent to two Heisenberg
spins $\gamma=\{a,b\}$ from the $i$th rung and $\sigma_{\gamma,j}=\pm1/2$
refer to two Ising spins $\gamma=\{a,b\}$ from the $j$th rung (see
Fig.~\ref{fig:Models}(c) for a schematic illustration). The exchange
constants $J_{0}$ and $J_{1}$ label the Ising intra-rung and intra-leg
interactions, while the XXZ Heisenberg intra-rung interaction is determined
by its $xy$-component $J_{x}$ and $z$-component $J_{z}$.

It has been proved in Ref. \cite{roj16} that the spin-1/2 Ising-XXZ
two-leg ladder can be rigorously mapped onto the mixed spin-3/2 and
spin-1/2 Ising-Heisenberg diamond chain, which can be subsequently
exactly solved within the transfer-matrix method. In a consequence
of that, one can obtain the exact expression for the free energy of
the spin-1/2 Ising-XXZ two-leg ladder (see Eq. (40) in Ref. \cite{roj16}),
which is formally identical with the formula \eqref{eq:free-energ-1}
of the effective Ising chain depending on three different Boltzmann's
factors. Owing to this fact, the spin-1/2 Ising-XXZ two-leg ladder
may display a similar quasitransition as the one-dimensional models
studied in the previous section whenever the three effective Boltzmann's
factors satisfy the condition \eqref{cc}.

\begin{figure}
\includegraphics[scale=0.22]{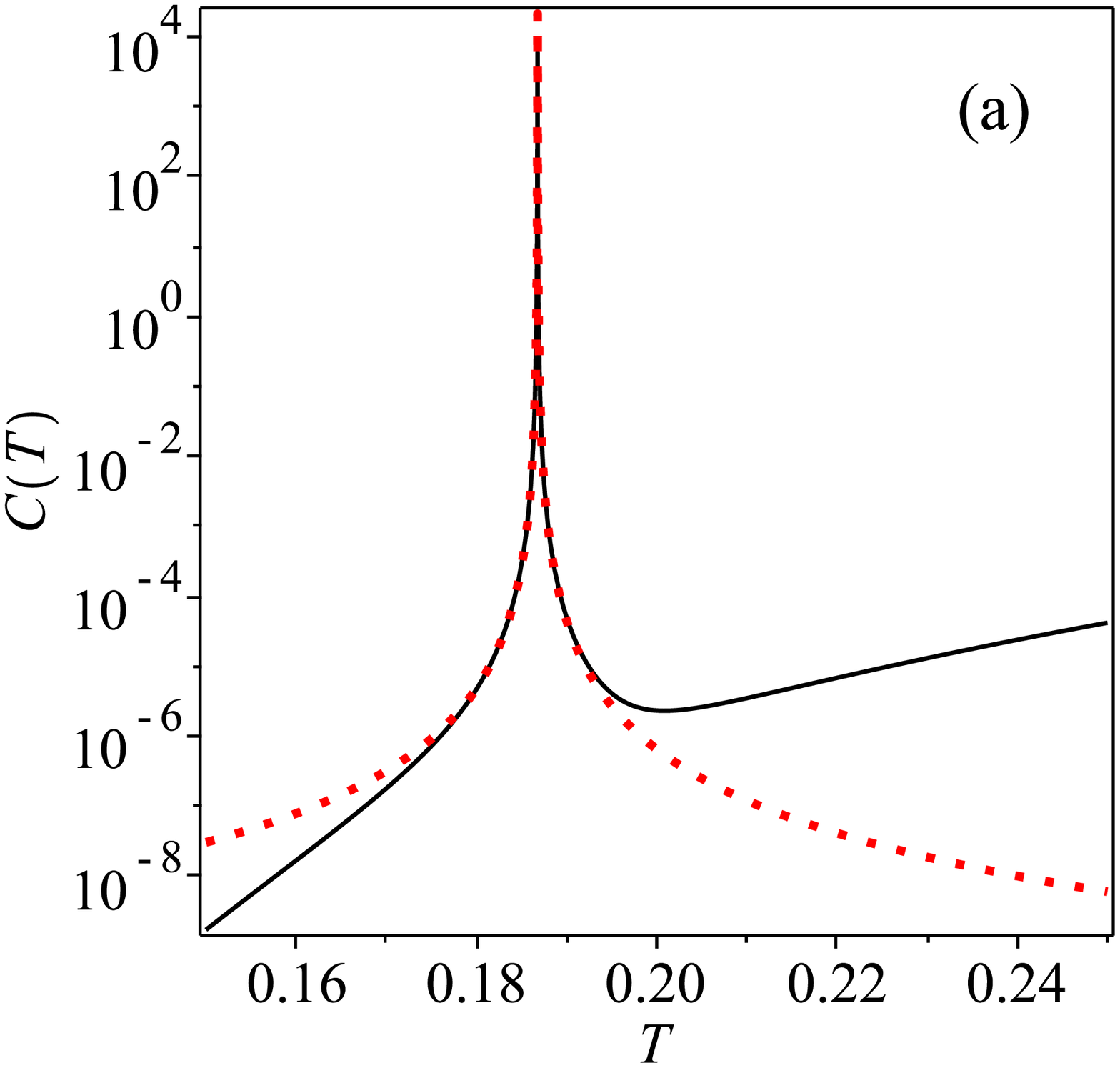}\includegraphics[scale=0.22]{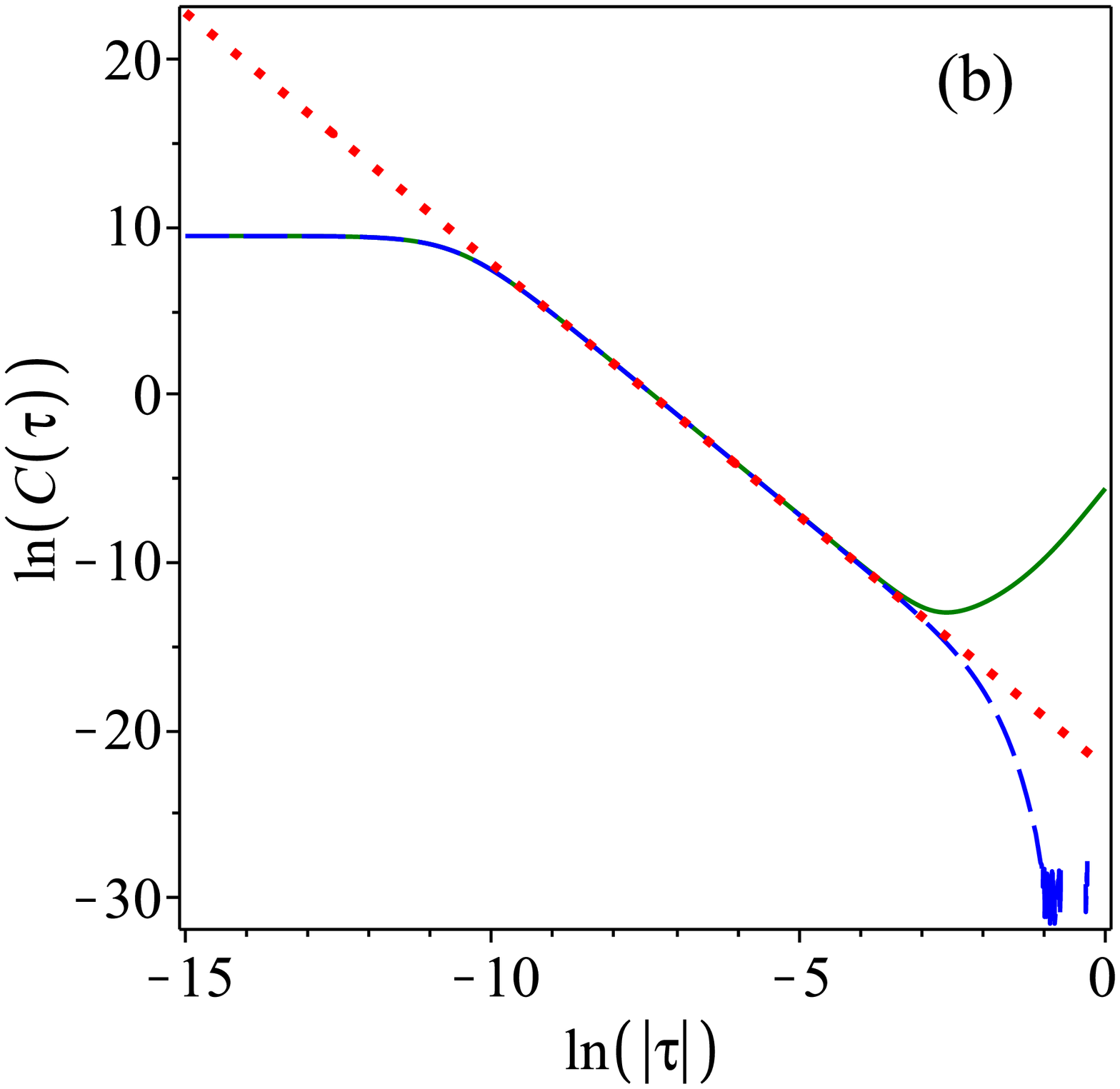}
\caption{\label{fig:laddelspc} Temperature variations of the specific heat
of the spin-$1/2$ Ising-XXZ two-leg ladder in a vicinity of the quasicritical
temperature by assuming the following set of coupling constants $J_{0}=25$,
$J_{x}=21.8$, $J_{z}=25$ and $J_{1}=-30$: (a) exact results (solid
line) for $C$ versus $T$ dependence is compared to the power-law
function (dotted line); (b) exact results for $\ln(C(\tau))-\ln(|\tau|)$
dependence above (below) the quasicritical temperature $\tau<0$
($\tau>0$) shown as a solid (dashed) line are compared to the power-law
function $\ln(C(\tau))=-3\ln(|\tau|)-22.2$ depicted by a dotted line.}
\end{figure}

To support this statement, the specific heat of the spin-1/2 Ising-XXZ
two-leg ladder is displayed in Fig.~\ref{fig:laddelspc}(a) a function
of temperature for the fixed values of the interaction constants $J_{0}=25$,
$J_{x}=21.8$, $J_{z}=25$ and $J_{1}=-30$ being responsible for
a quasitransition at the quasicritical temperature $T_{p}=0.186778$.
The solid line corresponds to the exact results derived according
to Ref. \cite{roj16}, while the dotted line denotes the relevant
power-law function. Temperature variations of the specific heat, which
are shown in Fig.~\ref{fig:laddelspc}(b) in the form of $\ln(C(\tau))$
versus $\ln(|\tau|)$ plot, bear evidence that the massive rise of
the specific heat sufficiently close but not too close to the quasicritical
temperature is driven by the power-law function $\ln(C(\tau))=-3\ln(|\tau|)-22.2$
being consistent with the quasicritical exponents $\alpha=\alpha'=3$.
From this perspective, the critical exponents of the spin-1/2 Ising-XXZ
two-leg ladder belong to the same universality class as reported previously
for the one-dimensional lattice-statistical models, which can be rigorously
mapped onto the effective Ising chain.

\subsection{Ising-XXZ three-leg tube}

Second, we will also investigate a quasitransition of the spin-$1/2$
Ising-XXZ three-leg tube \cite{str16} shown in Fig.~\ref{fig:Models}(d),
which takes into account the XXZ intra-triangle interaction between
the spins from the same triangular unit and the Ising inter-triangle
interaction between the spins from neighboring triangular units. The
Hamiltonian of the spin-$1/2$ Ising-XXZ three-leg tube is defined
as 
\begin{eqnarray}
{H}\!\!\! & = & \!\!\!\sum_{i=1}^{N}\sum_{j=1}^{3}\left[J_{x}\left({S}_{i,j}^{x}{S}_{i,j+1}^{x}+{S}_{i,j}^{y}{S}_{i,j+1}^{y}\right)+J_{z}{S}_{i,j}^{z}{S}_{i,j+1}^{z}\right]\nonumber \\
\!\!\! & + & \!\!\!J_{1}\sum_{i=1}^{N}\left(\sum_{j=1}^{3}{S}_{i,j}^{z}\right)\left(\sum_{j=1}^{3}{S}_{i+1,j}^{z}\right),\label{1}
\end{eqnarray}
where ${S}_{i,j}^{\alpha}$ $(\alpha\in\{x,y,z\})$ denote three spacial
components of the spin-$1/2$ operator, the first subscript $i$ specifies
a triangular unit in the three-leg tube and the second subscript $j$
determines a position of individual spin in a given triangular unit.
The coupling constants $J_{x}$ and $J_{z}$ denote the XXZ intra-triangle
interaction between the spins belonging to the same triangular unit,
while the other interaction term $J_{1}$ refers to the Ising inter-triangle
interaction between the spins from neighboring triangular units.

\begin{figure}
\includegraphics[scale=0.22]{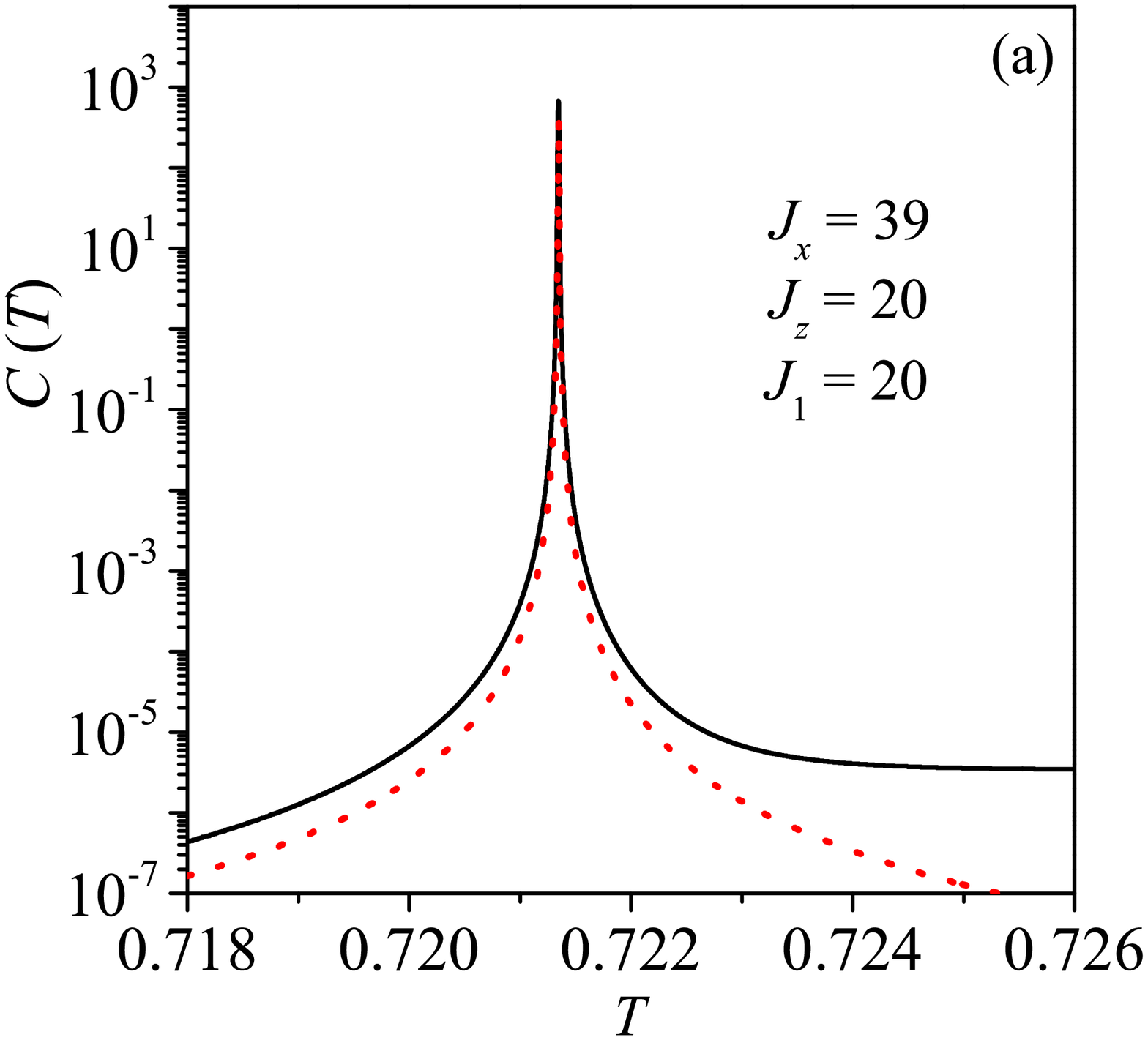}\includegraphics[scale=0.22]{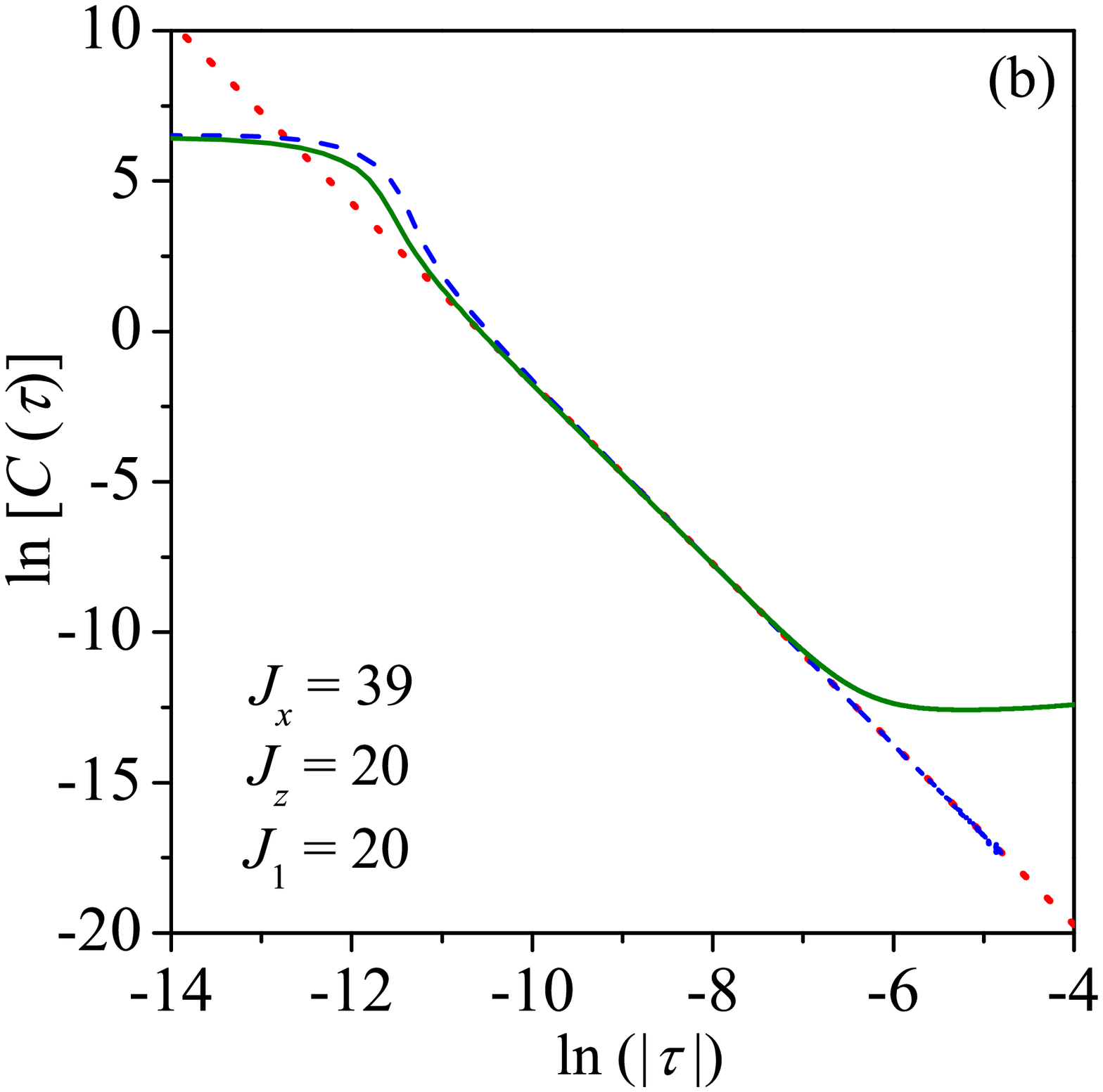}
\caption{\label{tube} Temperature variations of the specific heat of the spin-$1/2$
Ising-XXZ three-leg tube in a vicinity of the quasicritical temperature
by assuming the following set of coupling constants $J_{1}=20$, $J_{z}=20$
and $J_{x}=39$: (a) exact results (solid line) for $C$ versus $T$
dependence is compared to the power-law function (dotted line); (b)
exact results for $\ln(C(\tau))-\ln(|\tau|)$ dependence above (below)
the quasicritical temperature $\tau<0$ ($\tau>0$) shown as a solid
(dashed) line are compared to the power-law function $\ln(C(\tau))=-3\ln(|\tau|)-31.7$
depicted by a dotted line.}
\end{figure}

It is worthwhile to remark that the spin-$1/2$ Ising-XXZ three-leg
tube is fully quantum one-dimensional model because each spin of the
three-leg tube is involved in two XXZ exchange interactions and six
Ising interactions. In spite of this fact, the spin-$1/2$ Ising-XXZ
three-leg tube is still exactly solvable within the classical transfer-matrix
method because the total spin on a triangular unit represents locally
conserved quantity with well defined quantum spin numbers \cite{str16}.
The free energy and full thermodynamics of the spin-$1/2$ Ising-XXZ
three-leg tube has been reported in our previous work \cite{str16}
to which the readers interested in further details are referred to.
It is nevertheless worth noticing that the exact result for the free
energy of the spin-$1/2$ Ising-XXZ three-leg tube given by Eq. (12)
of Ref. \cite{str16} has similar structure as the formula \eqref{eq:free-energ-1}
of the effective Ising chain depending on three different Boltzmann's
factors.

In what follows, our attention will be limited to a detailed analysis
of a quasitransition of the spin-$1/2$ Ising-XXZ three-leg tube,
which is emergent at the following quasicritical temperature \cite{str16}
\begin{eqnarray}
T_{p}=\frac{4J_{1}-2J_{z}-J_{x}}{\ln4}.\label{peak}
\end{eqnarray}
For illustration, typical temperature variations of the specific heat
of the spin-$1/2$ Ising-XXZ three-leg tube are depicted in Fig.~\ref{tube}
by considering the set of interaction parameters $J_{1}=20$, $J_{z}=20$
and $J_{x}=39$, which give rise to a quasitransition at the quasicritical
temperature $T_{p}=1/\ln4\approx0.7213475$. Exact results for temperature
dependence of the specific heat $C(T)$ (solid line) derived according
to Ref. \cite{str16} indeed furnish evidence of the sizable peak,
which follows the power-law dependence $\ln(C(\tau))=-3\ln(|\tau|)-31.73$
if temperature is set sufficiently close but not too close to the
quasicritical temperature. This result would suggest that the same
quasicritical exponent $\alpha=\alpha'=3$ drives the relevant temperature
dependence of the specific heat of the spin-$1/2$ Ising-XXZ three-leg
tube near the quasicritical temperature. It might be therefore quite
reasonable to conjecture that there is just one unique set of quasicritical
exponents, which governs a quasitransition of one-dimensional lattice-statistical
models of very different nature.

It is worth to mention that the ladder model and three leg tube do
not consider the action of an external magnetic field. However, both
models still exhibit a quasitransition at zero field. 

We understand that, in the quasitransition, the system presents a
vigorous change in the local ordering on a strongly correlated scenario
but without showing a true symmetry breaking. Therefore, although
correlations change considerably during the quasicritical transition\cite{Isaac}
(with signatures in the response functions), there is no macroscopic
order parameter associated. Although the quasitransitions observed
at finite magnetic fields lead to a change in the sub-lattice magnetizations,
the sub-lattice magnetization remains null below and above the quasitransition
when it takes place at zero-field. 

\section{Conclusions}

In the present work, we have examined in detail the quasicritical
exponents of a general class one-dimensional lattice-statistical models
displaying a quasitransition at finite temperatures, which can be
rigorously solved through an exact mapping correspondence with the
effective Ising chain. The usefulness and validity of this approach
has been testified on two particular examples of exactly solved one-dimensional
models. In addition, the quasitransitions of other two one-dimensional
lattice-statistical models with short-range and non-singular interactions
were also dealt with. In any case the quasitransition of one-dimensional
models is characterized by intense sharp peaks in the specific heat,
magnetic susceptibility and correlation length, which are quite reminiscent
of divergences accompanying a continuous (second-order) phase transition.
It should be emphasized, however, that these intense sharp peaks are
always finite (even though of several orders of magnitude high) and
thus, they should not be confused with actual divergences accompanying
true phase transitions.

Despite of this fact, it has been verified that the sizable peaks
of the specific heat, magnetic susceptibility and correlation length
follow close to a quasitransition the power-law dependencies on assumption
that temperature is sufficiently close but not too close to the quasicritical
temperature. The quasicritical exponents of four paradigmatic exactly
solved lattice-statistical models, more specifically, the spin-1/2
Ising-XYZ diamond chain, the coupled spin-electron double-tetrahedral
chain, the spin-1/2 Ising-XXZ two-leg ladder and the spin-1/2 Ising-XXZ
three-leg tube, have turned out to be the same. Bearing all this in
mind, it appears worthwhile to conjecture a new universality class
for one-dimensional lattice-statistical models displaying a quasitransition
at finite temperatures, which is characterized by the unique set of
quasicritical exponents: $\alpha=\alpha'=3$ for the specific heat,
$\gamma=\gamma'=3$ for the susceptibility and $\nu=\nu'=1$ for the
correlation length. The conjectured values of quasicritical exponents
obviously violate the scaling relations satisfied at true phase transitions
and hence, they might be of benefit for experimentalists in distinguishing
true phase transitions from quasitransitions. A further test of this
universality hypothesis on other specific examples of one-dimensional
lattice-statistical models (e.g. fully classical Ising or Potts models,
fully quantum Heisenberg or Hubbard models, etc.) represents a challenging
task for future work.

Concerning experimental realization, it is noteworthy that the quasicritical
behavior is not specialty of one-dimensional Ising-Heisenberg spin
models, but according to our preliminary calculations, it may be also
found in several Ising spin chains and Heisenberg spin chains\cite{S-HW,Heuvel-ch,Bel-Oh,Sahoo,S-Honda,Han-Strecka,Torr-Jmmm}
significantly extending a class of one-dimensional magnetic compounds
for experimental testing. A more thorough analysis of pure Ising and
Heisenberg spin chains displaying quasicritical behavior will be subject
matter of future works. 
\begin{acknowledgments}
O. R. and S.M. de S. thank Brazilian agencies CNPq, FAPEMIG and CAPES.
M.L.L. thank the Alagoas state agency FAPEAL. J.S. acknowledges the
financial support by grant of The Ministry of Education, Science,
Research and Sport of the Slovak Republic under Contract No. VEGA 1/0531/19 and by grant of the Slovak Research and Development Agency
under Contract No.APVV-14-0073. 
\end{acknowledgments}

\appendix

\section{\label{appdxA}Alternative coefficient expression}

Alternatively, the coefficient \eqref{eq:coef-a1-a-1} can be expressed
using the eq.\eqref{eq:wnk}, here we assume only for convenience
$\varepsilon_{_{n,0}}$ as the lowest energy. Thus, we want to express
Boltzmann's factors around the quasicritical temperature. Then we
begin to manipulate the following expression 
\begin{alignat}{1}
\frac{w_{n}}{\tilde{w}_{n}}= & \frac{\sum\limits _{k=0}g_{n,k}{\rm e}^{-\beta\varepsilon_{_{n,k}}}}{\sum\limits _{k=0}g_{n,k}{\rm e}^{-\beta_{p}\varepsilon_{_{n,k}}}}\nonumber \\
= & \frac{{\rm e}^{-\beta\varepsilon_{_{n,0}}}\left\{ 1+\sum\limits _{k=1}\frac{g_{n,k}}{g_{n,0}}{\rm e}^{-\beta\delta_{_{n,k}}}\right\} }{{\rm e}^{-\beta_{p}\varepsilon_{_{n,0}}}\left\{ 1+\sum\limits _{k=1}\frac{g_{n,k}}{g_{n,0}}{\rm e}^{-\beta_{p}\delta_{_{n,k}}}\right\} },
\end{alignat}
where $\delta_{n,k}=\varepsilon_{n,k}-\varepsilon_{n,0}$ for $k\geqslant1$,
and $\beta_{p}=1/k_{B}T_{p}$ with $T_{p}$ being the quasicritical
temperature.

Using this notation we have, 
\begin{alignat}{1}
\frac{w_{n}}{\tilde{w}_{n}}= & {\rm e}^{-(\beta-\beta_{p})\varepsilon_{_{n,0}}}\frac{\left\{ 1+\sum\limits _{k=1}\frac{g_{n,k}}{g_{n,0}}{\rm e}^{-\beta\delta_{_{n,k}}}\right\} }{A_{n}},\label{eq:wnn}
\end{alignat}
where $A_{n}=1+\sum\limits _{k=1}\frac{g_{n,k}}{g_{n,0}}{\rm e}^{-\beta_{p}\delta_{_{n,k}}}$
with $\delta_{n,k}\geqslant0$.

Now, by writing \eqref{eq:wnn} in terms of $\tau$, it becomes 
\begin{alignat}{1}
\frac{w_{n}}{\tilde{w}_{n}}= & {\rm e}^{-\frac{\tau}{T}\varepsilon_{_{n,0}}}\frac{\left\{ 1+\sum\limits _{k=1}\frac{g_{n,k}}{g_{n,0}}{\rm e}^{-\beta_{p}\delta_{_{n,k}}}{\rm e}^{-\frac{\tau\delta_{n,k}}{T}}\right\} }{A_{n}}.\label{eq:wtau}
\end{alignat}

We are interested in analyzing \eqref{eq:wtau} in the limit $\tau\rightarrow0$.
Then we can use Taylor series expansion in \eqref{eq:wtau}, which
results in 
\begin{alignat}{1}
\frac{w_{n}}{\tilde{w}_{n}}= & (1-\frac{\tau}{T_{p}}\varepsilon_{_{n,0}})\left\{ 1-\frac{1}{A_{n}}\frac{{\rm d}A_{n}}{{\rm d}\beta}\Bigr|_{\beta=\beta_{p}}\frac{\tau}{T_{p}}\right\} +\mathcal{O}(\tau^{2}).\label{eq:w-s-tau}
\end{alignat}

Simplifying \eqref{eq:w-s-tau}, we have 
\begin{alignat}{1}
\frac{w_{n}}{\tilde{w}_{n}}= & 1-\frac{1}{T_{p}}\left(\varepsilon_{_{n,0}}+\frac{{\rm d}\ln(A_{n})}{{\rm d}\beta_{p}}\right)\tau+\mathcal{O}(\tau^{2}),\label{eq:w/tw}
\end{alignat}
where $\frac{{\rm d}\ln(A_{n})}{{\rm d}\beta_{p}}\equiv\frac{{\rm d}\ln(A_{n})}{{\rm d}\beta}\bigr|_{\beta=\beta_{p}}$.

Denoting the coefficient $a_{n}=-\frac{1}{T_{p}}\left(\varepsilon_{_{n,0}}+\frac{{\rm d}\ln(A_{n})}{{\rm d}\beta_{p}}\right)$
independent of $\tau$, we can rewrite \eqref{eq:w/tw} as follow
\begin{alignat}{1}
w_{n}= & \tilde{w}_{n}\left(1+a_{n}\tau\right)+\mathcal{O}(\tau^{2}).\label{eq:w1apx}
\end{alignat}

Now let us write ($w_{1}-w_{-1}$) using the relation \eqref{eq:w1apx},
so we obtain 
\begin{equation}
w_{1}-w_{-1}=\tilde{w}_{n}(a_{1}-a_{-1})\tau+\mathcal{O}(\tau^{2}).\label{eq:low-tau}
\end{equation}
We can write more explicitly $(a_{1}-a_{-1})$ as follow

\begin{alignat}{1}
a_{1}-a_{-1}= & \left[\varepsilon_{_{1,0}}-\varepsilon_{_{-1,0}}+\frac{{\rm d}\ln(\frac{A_{1}}{A_{-1}})}{{\rm d}\beta_{p}}\right].\label{eq:diff-a1s}
\end{alignat}

Here $a_{-1}$ and $a_{1}$, may depend of some parameter $x$, fixed
in quasicritical point by $x_{p}$, e.g. the external magnetic field
$h_{p}$.

\end{document}